\documentclass[aps,preprint,preprintnumbers,nofootinbib,showpacs]{revtex4}
\usepackage{epsfig}
\usepackage{color}
\usepackage{amsmath}
\usepackage{amssymb}
\usepackage{dsfont}
\usepackage[normalem]{ulem}
\usepackage{appendix}

\def\bea{\begin{eqnarray}}
\def\eea{\end{eqnarray}}
\def\sea{\nonumber \\&&}
\def\lla{\left\langle}
\def\rra{\right\rangle}
\def\za{\alpha}
\def\zb{\beta}
\def\zc{\gamma}
\def\ssc{\scriptscriptstyle}
\def\lsim{\mathrel{\raise.3ex\hbox{$<$\kern-.75em\lower1ex\hbox{$\sim$}}} }
\def\gsim{\mathrel{\raise.3ex\hbox{$>$\kern-.75em\lower1ex\hbox{$\sim$}}} }

\begin{document}

\preprint{{\vbox{\hbox{NCU-HEP-k078}
\hbox{rev. Dec 2020}
\hbox{ed. Sep 2021}
}}}

\title{Noncommutative Coordinate Picture of the Quantum Phase Space
\vspace*{.3in} 
}
\author{Otto C. W. Kong and Wei-Yin Liu}

\address{Department of Physics and 
Center for High Energy and High Field Physics,
National Central University, Chung-Li 32054, Taiwan
\vspace*{1.in}
}
%

\vspace*{.5in}
\begin{abstract}\vspace*{0.3in}
We illustrate an isomorphic representation of the observable
algebra for quantum mechanics in terms of the functions on 
the projective Hilbert space, and its Hilbert space analog, with 
a noncommutative product in terms of explicit coordinates and 
discuss the physical and dynamical picture. The isomorphism 
is then used as a base for the translation of the differential 
symplectic geometry of the infinite dimensional manifolds
onto the observable algebra as a noncommutative geometry. 
Hence, we obtain the latter from the physical theory itself.
We have essentially an extended formalism of the 
Sch\"rodinger versus Heisenberg picture which we describe 
mathematically as like a coordinate map from the phase 
space, for which we have presented argument to be seen 
as the quantum model of the physical space, to the 
noncommutative geometry coordinated by the six position 
and momentum operators. The observable algebra is taken
essentially as an algebra of formal functions on the latter 
operators. The work formulates the intuitive idea that the 
noncommutative geometry can be seen as an alternative, 
noncommutative coordinate, picture of familiar quantum phase 
space, at least so long as the symplectic geometry is concerned.
%
\end{abstract}
\maketitle

\section{Introduction}
In classical physics, physical quantities or observables are 
modeled by the real valued variables and physical states are
identified as points of the phase space which is a geometric 
structure modeled locally on a finite Cartesian product of
the real number lines. Such geometric space are called 
manifolds. In fact, a state can be described by the values of its 
coordinate variables, the position and momentum observables. 
These are the basic observables, combinations of which
(like as smooth functions)  essentially give all the others. 
From the mathematical point of view, there is a duality 
correspondence between the algebra of observable 
and the geometric structure. The algebra of observables of 
a classical theory is a commutative algebra, to be identified 
as functions on the corresponding commutative geometric 
space (the phase space),  a symplectic manifold. For a quantum 
theory the algebra of observables  is a noncommutative 
(operator) one.  In the spirit of noncommutative geometry,
one expects a dual geometric structure which is noncommutative.
Intuitively, one would naturally think of the latter as having 
noncommutative coordinate observables which can be given 
by the position and momentum operators. We seek a point 
of view based on physics to understand such a noncommutative 
geometric structure beyond real number manifolds, at least 
beyond the finite dimensional ones.

In the case of simple quantum mechanics, physicists however 
have a well established picture of the phase space as a symplectic
geometry. It is the Hilbert space, an infinite dimensional 
complex vector space, or the projective Hilbert space, an infinite 
dimensional, actually curved, manifold. The latter, as the manifold 
of the pure states \cite{S,AS,Cps}, is also a geometric structure 
dual to the noncommutative algebra of observables. Actually, for 
$C^*$-algebras, which are the class of algebras considered to be the 
proper setting for the mathematics of noncommutative geometry
 \cite{C} and the general idea of an observable algebra in a physical 
theory \cite{St,Em}, the space of pure states essentially always
has the more familiar commutative geometric structure of K\"ahler 
manifolds or K\"ahler bundles  \cite{Cps,kb}. Typically, they are 
infinite dimensional. A key perspective here is that a noncommutative 
observable/quantity can be modeled by an infinite number of 
commutative variables. A noncommutative coordinate in particular 
can be described as an infinite set of the real or complex 
coordinates. {\em The noncommutative geometry of the observable 
algebra for quantum mechanics} may be taken as nothing more than 
{\em an alternative or better, more intuitive, picture of the infinite 
dimensional symplectic geometry of the quantum phase space}. It 
may be somewhat similar to the intrinsic description of a curved 
manifold versus its extrinsic description as part of an Euclidean space.

The idea that a noncommutative observable has the information
content of an infinite number of real or complex numbers is easy
to appreciate, though it has not been taken seriously enough in 
our opinion. A quantum operator can be thought of as a matrix on 
the infinite dimensional Hilbert space characterized by the matrix 
elements in a chosen basis, {\em i.e.} a system of coordinates. 
In its eigenstate basis in particular, it is described by the set 
of eigenvalues.   The latter, though often considered, is not the 
most convenient description of the structure for the full algebra.  
{\em We have introduced the notion of a noncommutative
value of an observable which plays a key role in the explicit
identification of the six position and momentum operators 
as coordinates for the quantum phase space \cite{079,081}.} 
For a fixed state, the definite noncommutative value carries 
mathematically the full information about the observable
the theory contains. It is experimentally accessible, at least 
in principle.  

The (K\"ahlerian) geometrical picture of quantum mechanics has 
a slow development. Only after a major part of the last century 
we have a more comprehensive picture of it available, given in 
Ref.\cite{CMP}.  The paper also gives the extremely important 
result of an isomorphic description of the observable algebra 
as an algebra of the so-called  K\"ahlerian functions on the 
projective Hilbert space which generates Hamiltonian flows 
preserving the K\"ahler structure of the manifold, therefore 
also the metric. The metric is the one of a constant holomorphic 
sectional curvature fixed by the Planck constant $\hbar$. 
Therefore, quantum noncommutativity can be seen as a curvature. 
A quantum observable as an operator on the Hilbert space can 
be matched to the set of complex values of the corresponding 
K\"ahlerian function of all the points. We will present below all 
those features in terms of explicit coordinates to make them 
more easily appreciated even by the readers less accustomed 
to the use of more abstract mathematics. {\em We will use 
the isomorphic algebraic structure to look at the differential 
geometric structure for the observable algebra, in terms of the 
noncommutative coordinates, matching to those of the Hilbert 
or the projective Hilbert space. That is like defining the notions of 
noncommutative geometry from the physics of quantum mechanics.}
The work formulates the idea that the noncommutative geometry 
of the observable algebra for quantum mechanics can be seen as 
an alternative, noncommutative coordinate, picture of quantum 
phase space, at least so long as the symplectic geometry is concerned.
  
In the next two sections, we review the geometric pictures 
of quantum mechanics on the Hilbert space and the projective 
Hilbert space, mostly in terms of explicit coordinates. We intend 
to give an optimal formulation of the known results, paying  
due attention to the proper physical dimensions of the various
quantities. Our key references are Refs.\cite{CMP,Sch,Sch2}; other 
references consulted include 
Refs.\cite{H,H2,H3,Ca,Ca2,Ca3,B,B2,B3,It,spp,spp2,spp3,BZ,CJ,Mt,MS}. 
The presentation sets the background for the following sections. 
In Sec.\ref{sec4}, we present some details of the K\"ahlerian 
functions for  the noncommutative coordinate observables of 
$\hat{x}_i$ and $\hat{p}_i$ and their complex combinations 
${\za}_i$ and $\bar{\za}_i$ under a convenient choice of 
coordinates for the Hilbert space and the projective Hilbert 
space. The explicit identification of a point from the 
noncommutative values of the coordinates will be illustrated.
In Sec.\ref{sec5} starts the exploration of the noncommutative
differential geometric structure of the observable algebra in line 
with the above mentioned idea. This will then be extended further 
in Sec.\ref{sec6} by a kind of coordinate transformation/map 
between the infinite set of complex coordinates and the six 
noncommutative coordinates, which can be considered as
an extension of the familiar Schr\"odinger and Heisenberg 
picture correspondence. The two sections present a complete
noncommutative picture of the symplectic differential geometry 
of the quantum phase space. For background references on 
noncommutative geometry relevant to our formulation here, 
we note in particular Refs. \cite{C,M,CH,Ho,CHZ,GP,DM}. Our 
presentation of that is however to be seen as directly dictated 
by the physical theory. The last section concludes the paper.

\section{Geometric Particle Dynamics on the Hilbert Space \label{sec2}}
Let us first recall some basics of the symplecto-geometrical formulation of the
quantum mechanics. To start with, the Schr\"odinger equation, as the equation
of motion for a quantum state described by the vector $\left|\phi\rra$, can 
be casted into the form of the Hamiltonian equations of motion. Take an 
orthonormal basis $\left|n\rra$ for the Hilbert space $\mathcal{H}$
(of countably infinite dimension, ${{n}}=0$ to $\infty$), we have 
$\left|\phi\rra=\sum_n z^{{n}} \left|n\rra$  where the complex 
coordinates $z^{{n}}=\tilde{q}^{{n}}+i\tilde{p}^{{n}}$ of the state 
vector have (real coordinates) $\tilde{q}^{{n}}$ and $\tilde{p}^{{n}}$
satisfying
\bea &&
\frac{d \tilde{q}^{{m}}}{dt} = 
 \frac{\partial H(\tilde{q}^{{n}},\tilde{p}^{{n}})}{\partial \tilde{p}_{{m}}} \;,
\sea
\frac{d \tilde{p}^{{m}}}{dt} = -
 \frac{\partial H(\tilde{q}^{{n}},\tilde{p}^{{n}})}{\partial \tilde{q}_{{m}}} \;,
\eea
where the Hamiltonian function $H(\tilde{q}^{{n}},\tilde{p}^{{n}})$
is given by $\frac{1}{2\hbar}\lla\phi |\hat{H}  |\phi \rra$ for the Hamiltonian 
operator $\hat{H}$. Moreover, if we take $\left|n \rra$ to be the
eigenstates of $\hat{H}$ (assuming a discrete spectrum) with 
$\hat{H}\left|n \rra = \hbar \omega_{{n}} \left|n \rra$,
we have simply
\bea
H(\tilde{q}^{{n}},\tilde{p}^{{n}})
=\frac{1}{2\hbar} \lla\phi |\hat{H}  |\phi \rra
=\frac{1}{2}  \omega_{{n}} [ (\tilde{q}^{{n}})^2+(\tilde{p}^{{n}})^2]
\quad \mbox{(with summation)}
\eea
with 
\bea&&
\frac{d \tilde{q}^{{n}}}{dt} = \omega_{{n}} \tilde{p}^{{n}} \qquad \mbox{ (no summation)} \;,
\sea
\frac{d \tilde{p}^{{n}}}{dt} = - \omega_{{n}} \tilde{q}^{{n}} \qquad \mbox{(no summation)} \;.
\eea
Each of the configuration $\tilde{q}^{{n}}$ and the momentum
variables $\tilde{p}^{{n}}$ behaves exactly in the same way as 
those of a harmonic oscillator with frequency $\omega_{{n}}$ 
and the magnitude and phase of each complex coordinate 
$z^{{n}}$ serve as an action-angle variable pair of the 
completely integrable quantum system. The equations 
of motion are equivalent to
\bea &&
\frac{d {z}^{{m}}}{dt} = 
 -2i \frac{\partial H({z}^{{n}},\bar{{z}}^{{n}})}{\partial \bar{{z}}_{{m}}}
= -i  \omega_{{m}} {z}^{{m}}  \quad \mbox{ (no summation)} \;,
\sea
\frac{d  \bar{{z}}^{{m}}}{dt} = 
2i \frac{\partial H({z}^{{n}},\bar{{z}}^{{n}})}{\partial {z}_{{m}}} 
= i  \omega_{{m}} \bar{{z}}^{{m}}  \qquad \mbox{\ \ \  (no summation)} \;,
\eea
which are just the conjugates of each other.\footnote{
Note that with these coordinates, we always
have  $ {z}_{{m}}=\delta _{{m}{n}}  {z}^{{n}}$
and $\frac{\partial}{\partial {{z}_{{m}}}} = \delta^{{m}{n}}  
 \frac{\partial}{\partial {{z}^{{n}}}}$,
independently of the metric; writing the tangent vector 
$\partial_{{n}}= \frac{\partial}{\partial {{z}^{{n}}}}$,
the covector $\partial^{{n}}$ is metric dependent and cannot
be taken as $\frac{\partial}{\partial {{z}_{{n}}}}$.
} 
The analysis above illustrates a couple of basic things very explicitly. 
From dimensional analysis, the proper physical unit for the 
coordinates is $\sqrt{\hbar}$, which is the right unit for the position 
and momentum when expressed in the same unit.  For any choice
of $\hat{H}$ beyond the physical energy observable, the `Hamiltonian
equations of motion' are preserved under a scaling of all coordinates 
with any complex number, suggesting a description with the symmetry 
reduction. The latter is the formulation on the projective Hilbert space
given in the following section.  Note that the symmetry of a complex 
phase rotation of a state vector in particular illustrates the lack of  
independent meaning of the notion of configuration space and 
momentum space. We have argued that the correct perspective is 
for the phase rotation symmetry to be taken as a part of the fundamental 
(quantum) relativity symmetry for quantum mechanics, which says 
the quantum phase space is the proper model of the physical space 
at the quantum level \cite{066,070} .

The Hilbert space $\mathcal{H}$ can be taken as a K\"ahler manifold 
with a trivial metric 
$G_{{m} \bar{{n}}} =\frac{1}{2} \delta_{{m} {{n}}}$, 
and a symplectic form $\widetilde\omega_{{m} \bar{{n}}} 
           =i G_{{m} \bar{{n}}}$.
The tangent space of a vector space can be identified with itself.
$G$ and $\widetilde\omega$ correspond to the real and imaginary
part of the inner product, {\em i.e.}
\bea
\lla \psi| \phi \rra =  G(\left|\psi\rra,\left|\phi\rra) + i \,\widetilde\omega(\left|\psi\rra,\left|\phi\rra)\;,
\eea
with $G(\left|\psi\rra,\left|\phi\rra) = \widetilde\omega(\left|\psi\rra, i \left|\phi\rra)
= - \widetilde\omega(i\left|\psi\rra,  \left|\phi\rra)$.
The equations of motion have the standard form
\bea \label{zeq}
\frac{d {z}^{{m}}}{dt}
 = -2i \delta^{{m} {{n}}} \bar\partial_{{{n}}} H
= \widetilde\omega^{{m} \bar{{n}}} (dH)_{\bar{{n}}} 
=\widetilde{X}_{\!\ssc H}^{{m}}(z)\;,
\eea 
for the Hamiltonian function $H$ corresponding to the operator $\hat{H}$, 
where $\widetilde{X}_{\!\ssc H}$ is the Hamiltonian vector field. Note that 
$\widetilde\omega^{{m} \bar{{n}}}
 =-i G^{{m} \bar{{n}}}=-2i \delta^{{m} {{n}}}$. 
The above equation is just a geometrical/coordinate description 
of the action of the Schr\"odinger vector field 
$\widetilde{X}_{\!\ssc \hat{H}}=\frac{1}{i\hbar} \hat{H}$ 
on a state vector. Actually, we have for a tangent
vector $\left|Y\rra$ to $\mathcal{H}$ at $\left|\phi\rra$ 
\bea
d H(\left|\phi\rra) (Y) &=& \left.\frac{d}{dt} H(\left|\phi\rra+t \left|Y\rra)\right|_{t=0}
=\frac{1}{2\hbar} \left( \lla Y|\hat{H} |\phi \rra +\lla \phi | \hat{H} | Y\rra \right)
\sea \label{svf}
= G\!\left( \frac{1}{\hbar} \hat{H} \left|\phi \rra,  \left|Y\rra \right)
= \widetilde\omega(\widetilde{X}_{\!\ssc \hat{H}},Y) (\left|\phi\rra) \;,
\eea
for Hermitian $\hat{H}$. We have been only passing between the geometrical
language and the one of the operators and state vectors on $\mathcal{H}$. 
Each vector is a point of the space; an observable, as an operator, is completely 
characterized by its values on all possible states and should be seen as a function 
on the space as suggested by the symplectic formulation. How to properly 
think about those values and the issue of noncommutativity and Heisenberg 
uncertainties is a question we will address carefully below. For any Hermitian 
operator $\hat{K}$, we can introduce the Hamiltonian function for the symplectic
geometry, $K(\left|\phi\rra)=\frac{1}{2\hbar}\lla\phi |\hat{K}  |\phi \rra$.
The Poisson bracket between two functions $H$ and $K$ is given by
$\{H,K\}_{\ssc \widetilde\omega}=  \widetilde\omega(\widetilde{X}_{\!\ssc H},\widetilde{X}_{\!\ssc K})$,
defined and considered for smooth complex valued functions, though 
the Hamiltonian function for a (Hermitian) Hamiltonian operator is real. 
Extending the observable algebra to allow the complex linear combinations 
of Hermitian operators and consider it as having a $C^*$-algebra structure
is the generally adopted approach to the algebraic or noncommutative
geometric  formulation of quantum mechanics.  Such non-Hermitian
operators are not quite any less `observable' compared to the Hermitian
parts in the linear combination. A Hamiltonian function of such an
operator is a complex function the real and imaginary parts of which
are Hamiltonian functions for Hermitian operators. Notice though, not 
all Hamiltonian functions correspond to operators, Hermitian or otherwise, 
in the observable algebra. The functions that do  are called K\"ahlerian 
functions,  and have the Hamiltonian flows preserving the K\"ahler 
structure and therefore the metric, giving isometries.
Geometrically, we have 
\bea
\{H,K\}_{\ssc \widetilde\omega} \! = \! (dH)_{{m}} \widetilde\omega^{{m} \bar{{n}}} (dK)_{\bar{{n}}} 
  +  \!(dH)_{\bar{{m}}} \widetilde\omega^{\bar{{m}} {{n}}}  (dK)_{{n}} 
=\!  -2i  \! \left( \partial_{{m}} H   \delta^{{m}{n}} \bar\partial_{{n}} K
    -\partial_{{m}}  K  \delta^{{m}{n}}  \bar\partial_{{n}} H \right) ,
\eea
where $\bar{\partial}_{{m}} \equiv \partial_{\bar{{m}}}
 =\frac{\partial}{\partial z^{\bar{{m}}}}= \frac{\partial}{\partial \bar{z}^{{{m}}}}$.
Using the coordinates and the matrix elements of the operators,
one can easily obtain
\bea \label{kp}
  \partial_{{m}}  H  \,G^{{m} \bar{{n}}} \partial_{\bar{{n}}} K
   \, (\left|\phi\rra) 
= \frac{1}{2\hbar^2}  \lla \phi | \hat{H}\hat{K} | \phi \rra \;.
\eea
This is a very simple but remarkable result, the key result behind 
the whole analysis. The first application of it gives
\bea \label{pb}
\frac{d}{dt} K (\left|\phi\rra) 
= \{K,H\}_{\ssc \widetilde\omega} (\left|\phi\rra) 
 = \frac{1}{2i\hbar^2}  \lla \phi | [\hat{K},  \hat{H}] | \phi \rra \;.
\eea
The latter is equivalent to the Heisenberg equation of motion 
under the Hamiltonian $\hat{H}$, namely 
$\frac{d}{dt} \hat{K}  =   \frac{1}{i\hbar}  [\hat{K}, \hat{H}]$.
The symplecto-geometrical form, however, works also for the
complex functions, which suggests to include the non-Hermitian 
operators in the Heisenberg equation of motion. In terms of the 
corresponding Schr\"odinger vector fields (for Hermitian operators), 
one can get the same equation as
\bea
\{K,H\}_{\ssc \widetilde\omega} (\left|\phi\rra) 
  = \widetilde\omega\!\left( \frac{1}{i\hbar} \hat{K}, \frac{1}{i\hbar} \hat{H} \right) (\left|\phi\rra)
=    \frac{1}{\hbar^2} \frac{1}{2i}  
 \left( \lla \hat{K}\phi |  \hat{H}  \phi \rra -  \lla \hat{H}\phi |  \hat{K}  \phi \rra \right)\;.
\eea
Notice that the Schr\"odinger vector field expressions from Eq.(\ref{svf})
is valid only for Hermitian operators. 
Denote the Hamiltonian function for an operator 
$\zb(\hat{p}_i,\hat{x}_i)$, for the three position and three momentum 
operators with $[\hat{x}_i,\hat{p}_j]=i\hbar \delta_{ij}$, by $H_{\!\ssc\zb}$ 
and the Hamiltonian  vector field by $\widetilde{X}_{\!\ssc\zb}$. Note that real 
functions correspond to Hermitian operators and the Hermitian
conjugate of $\zb$ is otherwise given by $\bar\zb$. Moreover
$H_{\!\ssc\bar\zb}= \bar{H}_{\!\ssc\zb}$.
The last part of Eq.(\ref{pb}) can be written as
\bea\label{pbz}
H_{\!\ssc[\zb,\zc]} 
=  \frac{1}{2\hbar}  \lla \phi | [\zb,\zc] | \phi \rra
= {i\hbar} \{H_{\!\ssc\zb},H_{\!\ssc\zc}\}_{\ssc \widetilde\omega}  \;.
\eea

While a generic symplectic manifold may not possess a Riemannian metric, 
with the K\"ahler structure, however, we have the latter being intimately 
connected to the symplectic structure. In fact, a symplectic form together
with a compatible complex structure on a manifold uniquely fixes the metric.
We have seen that the Poisson algebra of $H_{\!\ssc\zb}$ functions gives 
an isomorphic description of the Poisson algebra of operators $\zb$ with 
the commutator, multiplied by $\frac{1}{i\hbar}$, taken as the Poisson 
bracket. Our derivation here starts with the remarkable result of 
Eq.(\ref{kp}), expressed in terms of the metric.  The result actually gives 
a full isomorphism between the algebra of $H_{\!\ssc\zb}$ functions 
and the observable algebra. 

One can define the Riemann bracket
\bea \label{PRH}
\{ H_{\!\ssc\zb},H_{\!\ssc\zc} \}_{\!\ssc G} (\left|\phi\rra) 
 :=  G(\widetilde{X}_{\!\ssc H_{\!\ssc\zb}},\widetilde{X}_{\!\ssc H_{\!\zc}}) (\left|\phi\rra) 
= \frac{1}{2\hbar^2}  \;  \lla \phi | [\zb,\zc]_+| \phi \rra 
= \frac{1}{\hbar} H_{\!\ssc [\zb,\zc]_+}
\eea
in terms of the anticommutator. The latter can be seen as
the Riemann bracket for the observable algebra. Note that
the Jordan algebraic product,  called Jordan bracket,
is exactly half the anticommutator. We write
\bea
{(\zb,\zc)}_{\!\ssc J} = \frac{1}{2} [\zb,\zc]_+ \;.
\eea
It is exactly  the nonassociative Jordan 
product for the operators which is half the anticommutator.  
Furthermore, one can write the so-called K\"ahler product 
on the space of $H_{\!\ssc\zb}$ functions  in the simple form
\bea
H_{\!\ssc\zb} \star_{\!\ssc K} H_{\!\ssc\zc} = H_{\!\ssc\zb\zc} \;;
\eea
that is to say, the  K\"ahler product given in terms of $\star_{\!\ssc K} $ 
matches the structure of the operator product as the basic product 
between observables\footnote{
We have the even more suggestive form
$H_{\!\ssc\zb} \star_{\!\ssc K} H_{\!\ssc\zc} = H_{\!\ssc\zb\star\zc}$ for a
formulation of the observable algebra as the Moyal star
product algebra of functions of the real variables $\zb(p_i,x_i)$
and $\zc(p_i,x_i)$ \cite{070}.
}.
It is exactly given by Eq.(\ref{kp}), {\em i.e.} 
 $\hbar\, \partial_{{m}}  H_\zb  \,G^{{m} \bar{{n}}} \partial_{\bar{{n}}} H_\zc$.
Obviously, we have 
\bea\label{Hkp}
H_{\!\ssc\zb} \star_{\!\ssc K} H_{\!\ssc\zc} 
 =  \frac{\hbar}{2} \{ H_{\!\ssc\zb},H_{\!\ssc\zc} \}_{\ssc G}  +
  \frac{i\hbar}{2}    \{H_{\!\ssc\zb},H_{\!\ssc\zc}\}_{\ssc \widetilde\omega}  
  =  H_{\!\ssc{(\zb,\zc)}_{\!\ssc J}} + \frac{1}{2}  H_{\!\ssc [\zb,\zc]} 
=\frac{ H_{\!\ssc[\zb,\zc]_+} +  H_{\!\ssc [\zb,\zc]}}{2}\;,
\eea
as ${\zb\zc} =   {(\zb,\zc)}_{\!\ssc J}  +\frac{1}{2} [\zb,\zc]$, 
which is just the splitting of the operator product into 
the symmetric and antisymmetric parts.

\section{Geometric Particle Dynamics on the Projective Hilbert Space\label{sec3}}
The linearity of the Schr\"odinger equation says that all the state vectors 
differing by a nonzero constant factor behave in exactly the same way. The 
zero vector, however, does not correspond to a sensible physical state. This 
suggests a symmetry reduction of the symplectic system to one of one 
lower complex dimension, to the projective Hilbert space $\mathcal{P}$ 
with each ray of vectors $[\phi]$ identified as a point.  The latter space is  
an infinite dimensional complex projective space ($\mathds{CP}^\infty$), 
still a K\"ahler manifold. As the space of pure states, it is a geometric 
structure dual to the noncommutative $C^*$-algebra  as the algebra of 
observables. Any set of $z^{{n}}$ serves as a set of  homogeneous 
coordinates.  Natural atlas of affine coordinates is given in the form 
$w^{\tilde{n}}=\frac{z^{\tilde{n}}}{z^{\ssc 0}}$ with $\tilde{n}$  
counting from 1 to $\infty$.  Points corresponding to $[\phi]$ with 
vanishing $z^{\ssc 0}$ all have $w^{\tilde{n}}$ as infinity, although 
$\mathcal{P}$ is actually compact. If fact, one only has to switch 
to the another similar coordinate chart, for example one obtained 
by a swapping the $z^{{n}}$ coordinates first, to give such points 
finite coordinate values. \{$z^{{n}}$\}  as a system of coordinates 
on $\mathcal{P}$ with redundancy has the benefit of being globally 
applicable. Besides, the Hilbert space picture of quantum mechanics
is more than a convenient redundant description. Mathematically,
it is the natural structure to arrive at from the point of view of the 
representation theory of the observable algebra, or that of the
fundamental symmetry behind, which can or should be identified 
as the (quantum) relativity symmetry \cite{066,070}. Physically, 
the notion of the Berry's phase clearly indicates that there are 
nontrivial dynamical issues involving (the changes in) the $\theta$ 
coordinate for an overall phase factor \cite{CJ,Mt} which cannot 
be described on $\mathcal{P}$ alone.

From the geometry of the complex projective spaces $\mathcal{P}$, 
we have  the standard Fubini-Study metric given by
\bea \label{dsw}
ds^2 =2  \, g_{\tilde{m}\bar{\tilde{n}}} \, dw^{\tilde{m}} d\bar{w}^{\tilde{n}} 
 = \frac{2\hbar}{1+|w|^2} \left( \delta_{\tilde{m}\bar{\tilde{n}}}
  -\frac{\bar{w}_{\tilde{m}}w_{\tilde{n}}}{1+|w|^2}  \right) dw^{\tilde{m}} d\bar{w}^{\tilde{n}}\;.
\eea
with $|w|^2=\bar{w}_{\tilde{n}}w^{\tilde{n}}$. Note that 
we have supplemented the mathematical result by an $\hbar$ 
to keep the right physical dimension for $ds^2$, since $w^{\tilde{n}}$,
unlike $z^{{n}}$, has no length dimension, and 
adopted the factor of 2, which fits in with the physics 
results  presented below most nicely. The symplectic 
form $\omega$ is given by the K\"ahler form, with 
$\omega_{\tilde{m}\bar{\tilde{n}}}=i g_{\tilde{m}\bar{\tilde{n}}}$. 
We have also the inverse
\bea
g^{\bar{\tilde{n}}\tilde{m}} =\frac{1}{\hbar} ({1+|w|^2}) 
 \left( \delta^{\bar{\tilde{n}}\tilde{m}} + \bar{w}^{\tilde{n}} w^{\tilde{m}}  \right) \;.
\eea
In terms of the \{$z^{{n}}$\} set of the homogeneous coordinates, 
we can write the Fubini-Study metric as
\bea \label{dsz}
ds^2 = 2 \tilde{g}_{{m}{\bar{{n}}}}  \, dz^{{m}} d\bar{z}^{{n}} 
=   \frac{2\hbar}{|z|^2}    \left(   \delta_{{m}{n}} 
        - \frac{z_{{n}} \bar{z}_{{m}}}{|z|^2 } \right)
           dz^{{m}} d\bar{z}^{{n}}\;.
\eea
Note that $\det\!\tilde{g}=0$; the metric is hence formally degenerate. 
Of course $g_{\tilde{m}\bar{\tilde{n}}}$ is in itself not degenerate.
One can describe a point in $\mathcal{P}$ as the equivalent class 
$[\phi]$ of the Hilbert space vectors $\left|\phi\rra$, each being a
constant multiple of the others. We have
\bea \label{dsp}
ds^2 
 = 2\hbar \frac{\lla \delta\phi |\delta\phi \rra}{\lla \phi |\phi \rra}
  - 2\hbar \frac{\lla \delta\phi |\phi \rra \lla \phi |\delta\phi \rra}{\lla \phi |\phi \rra^2} \;, 
\eea
which corresponds to a distance between the two state vectors as given by
\bea
s(\left|\phi \rra, \left|\phi' \rra)=\sqrt{2\hbar} \cos^{-1}\!\!\sqrt\frac{|\!\!\lla \phi|\phi'\rra\!\!|^2}{ |\!\!\lla \phi'|\phi'\rra\!\!| |\!\!\lla \phi|\phi\rra\!\!|} \;.
\eea
It depends, of course, only on $[\phi]$ and $[\phi']$ and is the geodesic 
distance between the two points in $\mathcal{P}$ as the quantum 
model of the physical space. In the conventional picture of quantum
mechanics, it characterizes the distinguishability of the physical
states. The maximum value of $s$ is given by 
$\pi \sqrt{\frac{\hbar}{2}}$, realized between any two 
orthogonal state vectors.  

The metric in Eq.(\ref{dsz}) is exactly that of Eq.(\ref{dsp}) expressed in 
terms of the coordinates of $\mathcal{H}$, and the one in Eq.(\ref{dsw}) 
in terms of the affine coordinates of $\mathcal{P}$. One may then think 
of $\mathcal{H}-\{0\}$ as a complex line bundle over the base space 
$\mathcal{P}$ and of the degenerate metric as one for the whole bundle 
which vanishes on the vertical tangent vectors \cite{Mt}. Hence, the 
distance between the points within the same fiber is always zero. The 
idea that each $[\phi]$, rather than an individual $\phi$, represents 
a physical state suggests that the Fubini-Study metric, rather than the 
trivial Hilbert space metric, is the proper metric for the distance 
between physical states, or state vectors, in quantum mechanics. The 
$\{r, \theta,w^{\tilde{n}}, \bar{w}^{\tilde{n}}\}$ 
set with $r= |z|$ and $z^{\ssc 0}=re^{i\theta}(1+|w|^2)^{-1/2}$, as a 
full set of coordinates for $\mathcal{H}$ may be used best to illustrate 
that \cite{CJ}. The $\theta$ coordinate is an overall phase factor for 
a state vector and $r$ its magnitude. A transformation in the $\theta$ 
coordinate maintains the inner product of two vectors, hence also the 
symplectic form and the metric. The coordinate is therefore somewhat 
redundant or irrelevant to the geometric structure as well as to the 
dynamics. The identity operator as a `Hamiltonian' operator generates 
exactly a change in all state vectors by a `translation' of their $\theta$ 
coordinates by the value $-\frac{i}{\hbar}t$ producing the circle action of 
the group of $\theta$ transformations. $H_{\!\ssc 1}=\frac{r^2}{2\hbar}$ 
is the corresponding Hamiltonian function.  We have a standard 
case of symmetry reduction of a circle action  with $\mathcal{P}$ 
being isomorphic to the quotient of any regular level set of nonzero 
value of $H_{\!\ssc 1}$ and the circle $S^1$, or $(\mathcal{P}, \omega)$ 
the corresponding symplectic quotient of $(\mathcal{H}, \widetilde\omega)$ 
at the  constant $r$ \cite{MS}.  That is to say, all the Hamiltonian 
flows generated by  Hermitian operators stay on spheres of fixed 
radius $r$, with the dynamical picture on the each 
sphere of nonzero radius essentially identical. Note that all 
$H_{\!\ssc\zb}$ are $\theta$-independent. They actually have the 
form $r^2$ times a function of $w^{\tilde{n}}$ and $\bar{w}^{\tilde{n}}$. 
The latter is essentially the reduced Hamiltonian function on 
$\mathcal{P}$ which is the focus of this section.

Going back to  the metric structure, the trivial metric of 
$\mathcal{H}$ can be written in the form 
\[
ds^2_{\!\ssc (\mathcal{H})} =dr^2 + \frac{r^2}{2\hbar} ds^2_{\!\ssc (\mathcal{S})} \;,
\]
where $ds^2_{\!\ssc (\mathcal{S})}$ is the metric on the sphere 
at $r^2=2\hbar$, or the $r$-independent part of the full metric. 
The  $2\hbar$ factor is essentially the same one as in the 
case of the Fubini-Study metric.  The vector field $\partial_r$ 
is vertical, or orthogonal to the sphere. The metric tensor on 
$\mathcal{S}$ is hence given by 
$\frac{2\hbar}{r^2} ( G - \partial_r\otimes \partial_r )$.
In addition, the invariance of the metric with respect to the 
$\theta$-transformation gives a Killing reduction to the
$\theta$-independent metric \cite{Sch,Sch2,g71}, one 
on $\mathcal{P}$. The result is
\[
\frac{2\hbar}{r^2} ( G - \partial_r\otimes \partial_r ) 
- \frac{2\hbar}{r^4} \partial_\theta\otimes \partial_\theta \;,
\]
which is exactly the metric given in Eq.(\ref{dsz}) upon substituting 
$\partial_r = \frac{z^{{m}}}{r} \partial_{{m}}
  + \frac{\bar{z}^{{m}}}{r} \partial_{\bar{{m}}}$
and $\partial_\theta = iz^{{m}} \partial_{{m}}
  -i\bar{z}^{{m}} \partial_{\bar{{m}}}$. 
One can also take ${r} (\partial_r)$ together with $\partial_\theta$ 
to be the Killing vector fields for the conformal metric 
$\tilde{G}_{{m} {\bar{n}}} =\frac{2\hbar}{r^2}  G_{{m} {\bar{n}}}$ 
on $\mathcal{H}-\{0\}$ for a direct reduction and obtain the 
same result. The procedure is a powerful one, allowing us to get 
the other corresponding tensors on $\mathcal{P}$, including the 
useful `inverse metric', which can be obtained as
\bea \label{g-1z}
\tilde{g}^{m{\bar{n}}}  
=    \tilde{G}^{m{\bar{n}}}  -\frac{1}{2\hbar} r^2 \partial_r^{\,m} \partial_r^{\,\bar{n}}
 -\frac{1}{2\hbar} \partial_\theta^{\,m} \partial_\theta^{\,\bar{n}}
=      \frac{1}{\hbar} \left( |z|^2\delta^{m\bar{n}}   -  {z}^{{m}} \bar{z}^{{n}} \right) \;.
\eea
Note that the singular or degenerate metric $\tilde{g}_{m{\bar{n}}}$  
in Eq.(\ref{dsz}) cannot be inverted. We will apply the procedure
extensively in our analysis. More details on the involved mathematics 
are given in the Appendix, for readers' convenience. 

The Fubini-Study metric on $\mathcal{P}$, besides having the similar
role  as the metric on $\mathcal{H}$ for defining a K\"ahler product 
among functions representing the operators as observables,
has also an important role to play in relation to the  quantum 
covariance or the Heisenberg uncertainty, as illustrated below. 

Functions on $\mathcal{P}$ can be defined in terms of functions 
on $\mathcal{H}$ which are independent of the $r$ and $\theta$
coordinates. In particular, we consider the so-called K\"ahlerian 
functions on $\mathcal{P}$ given by
\bea
f_{\!\ssc\zb} ([\phi]) = 2 \hbar \frac{H_{\!\ssc\zb} (\left| \phi \rra)}{\lla \phi | \phi \rra} 
  = \frac{\lla \phi |\zb|\phi \rra }{\lla \phi | \phi \rra}  \;, 
\eea
each corresponding to the function of the expectation values 
for the operator $\zb$. Note that we introduced the $\hbar$ 
factor in the above definition so that $f_{\!\ssc\zb}$ has the 
same physical dimension as $H_{\!\ssc\zb}$ and the operator 
$\zb$ or its matrix elements.  The factor 2 is just for the convenience 
of having the corresponding functions of the constant operators
as the same constants. An extra constant factor in the definition
of $f_{\!\ssc\zb}$ otherwise does not really matter and it may
seem to be  natural to omit the factor of two so that $f_{\!\ssc\zb}$ 
agrees with $H_{\!\ssc\zb}$ for a normalized $\phi$ (up to $\hbar$).  
Our choices of exact forms of the $H_{\!\ssc\zb}$ and 
$f_{\!\ssc\zb}$ functions result in a $\frac{r^2}{2\hbar}$ 
factor difference, {\em i.e.} $r^2=2\hbar$ is where the two 
functions have the same value, and the $2\hbar$ factor 
shows up as the natural way to obtain the various metrics 
discussed above in the preferred form. Using either the 
$w^{\tilde{n}}$ coordinates and $g^{{\tilde{m}}\bar{{\tilde{n}}}}$ 
or the $z^n$ with $\tilde{g}^{m\bar{n}}$, we can easily see 
that for the K\"ahler product is defined by
\bea
 f_{\!\zb} \star_{\kappa} f_{\!\zc} 
 =   f_{\!\ssc\zb} f_{\!\ssc\zc}   + {\hbar}\,\partial_{\tilde{m}} f_{\!\ssc\zb} \, 
   g^{\tilde{m} \bar{\tilde{n}}} \partial_{\bar{\tilde{n}}} f_{\!\ssc\zc} 
=   f_{\!\ssc\zb} f_{\!\ssc\zc} 
   + {\hbar}\,\partial_m f_{\!\ssc\zb} \,   \tilde{g}^{m \bar n }\partial_{\bar n} f_{\!\ssc\zc} \;,
\eea
to give
\bea
 f_{\!\ssc\zb} \star_{\!\kappa} f_{\!\ssc\zc} = f_{\!\ssc\zb\zc} \;.
\eea
Again, from the antisymmetric and symmetric parts,
we have the Poisson and Riemann brackets 
\bea &&
\{ f_{\!\ssc\zb},f_{\!\ssc\zc} \}_{\tilde\omega} 
=\{ f_{\!\ssc\zb},f_{\!\ssc\zc} \}_\omega  
  = \frac{1}{i\hbar} f_{\!\ssc [\zb,\zc]} \;,
\sea\label{PRf}
\{ f_{\!\ssc\zb},f_{\!\ssc\zc} \}_{\tilde{g}}
=\{ f_{\!\ssc\zb},f_{\!\ssc\zc} \}_g
 =  \frac{1}{\hbar} f_{\!\ssc [\zb,\zc]_+} -\frac{2}{\hbar} f_{\!\ssc\zb} f_{\!\ssc\zc}
\eea
with 
\bea
 f_{\!\ssc\zb} \star_{\kappa} f_{\!\ssc\zc} 
 =    f_{\!\ssc\zb} f_{\!\ssc\zc} + \frac{\hbar}{2} \{ f_{\!\ssc\zb},f_{\!\ssc\zc} \}_g
     +\frac{i\hbar}{2} \{ f_{\!\ssc\zb},f_{\!\ssc\zc} \}_\omega  
= \frac{   f_{\!\ssc [\zb,\zc]_+} +  f_{\!\ssc [\zb,\zc]} }{2} 
\eea
(and the Jordan bracket
$f_{\!\ssc{(\zb,\zc)}_{\!\ssc J}} = \frac{1}{2}  f_{\!\ssc [\zb,\zc]_+}$).
In addition, we have
$\{ f_{\!\ssc\zb},f_{\!\ssc\zb} \}_g = \frac{2}{\hbar}  (\Delta\zb)^2$,
from the Heisenberg uncertainty
$(\Delta\zb)^2=\frac{\lla\phi|(\zb)^2|\phi\rra} {\lla\phi|\phi\rra}
   - \left(\frac{\lla\phi|\zb|\phi\rra}{\lla\phi|\phi\rra} \right)^2$
for the operator $\zb$. More directly, one can introduce the 
quantum covariance between the two operators as 
\[
\text{Cov}(\zb,\zc)  = 
  \lla \left( \zb - \lla \zb\rra,\zc - \lla \zc\rra \right)_{\!\ssc J} \rra
= \frac{\hbar}{2} \{ f_{\!\ssc\zb},f_{\!\ssc\zc} \}_g \;,
\]
in relation to which we have the inequality
\[
(\Delta\zb)^2 (\Delta\zc)^2 \geq 
  (\frac{\hbar}{2} \{ f_{\!\ssc\zb},f_{\!\ssc\zc} \}_{\omega})^2 
 + (\frac{\hbar}{2} \{ f_{\!\ssc\zb},f_{\!\ssc\zc} \}_g)^2 \;,
\]
as a stronger version of the Heisenberg uncertainty principle. 
All this illustrates the key role of the Riemann bracket or the 
metric on $\mathcal{P}$ in quantum mechanics compared to 
which the metric $G$ on $\mathcal{H}$ is much less 
physically relevant.

We see here another isomorphic description of the observable
algebra, namely the $f_{\!\ssc\zb}$ functions with the K\"ahler
product. To compare results here with those of the $H_{\!\ssc\zb}$
functions with  $H_{\!\ssc\zb}= \frac{r^2}{2\hbar} f_{\!\ssc\zb}$, 
the first point to note is that
\bea \label{pbe}
\{H_{\!\ssc\zb}, H_{\!\ssc\zc} \}_{\ssc \widetilde\omega} 
 = \frac{1}{i\hbar}  \frac{r^2}{2\hbar}  f_{\!\ssc [\zb,\zc]} 
=    \frac{r^2}{2\hbar}  \{ f_{\!\ssc\zb},f_{\!\ssc\zc} \}_{\tilde\omega}
 =    \frac{r^2}{2\hbar}  \{ f_{\!\ssc\zb},f_{\!\ssc\zc} \}_\omega  \;,
\eea
where we have exact equality of the two Poisson brackets 
for $r^2=|z|^2= 2\hbar$.  This is in line with the view
of  $\mathcal{P}$ as the symplectic reduction of
$\mathcal{H}$. In fact, applying explicitly the coordinate
transformation  from $(z^{{n}}, \bar{z}^{{n}})$ to
$(r,\theta,{w}^{\tilde{n}},\bar{w}^{\tilde{n}})$ 
onto the Hamiltonian vector field $\widetilde{X}_{\!\ssc H_{\!\ssc\zb}} (\equiv \widetilde{X}_{\!\ssc\zb})$ 
gives
\bea
\widetilde{X}_{\!\ssc H_{\!\ssc\zb}} 
  = \frac{r^2}{2\hbar} \widetilde{X}_{\!\ssc f_{\!\ssc\zb}} + \frac{f_{\!\ssc\zb}}{2\hbar} \widetilde{X}_{r^2}
=\tilde{X}_{\!\ssc\zb} + \frac{f_{\!\ssc\zb}}{2\hbar} \widetilde{X}_{r^2}
  = X_{\!\ssc\zb} +   \tilde{X}_{\!\ssc f_{\!\ssc\zb}}^\theta \partial_{\theta} 
            + \frac{f_{\!\ssc\zb}}{2\hbar} \widetilde{X}_{r^2} \;, 
\label{XHf}\eea
with $\tilde{X}_{\!\ssc\zb}$ being the horizontal lift of 
$X_{\!\ssc\zb}$, from the perspectives of the Killing reduction
 \cite{g71} or sub-Riemannian structure \cite{Mt} as the
relation between $\mathcal{P}$ and $\mathcal{H}-\{0\}$. 
(More results on the Killing reduction are presented in the Appendix.)
We have
\bea
\widetilde{X}_{r^2} = \widetilde{X}_{\!\ssc H_{\!\ssc 2\hbar}}= -2 \partial_\theta 
\eea
(the Hamiltonian vector field for the operator $2\hbar {I}$),
and $\widetilde{X}_{\!\ssc f_{\!\zb}}$ is the Hamiltonian 
vector field of $f_{\!\ssc\zb}$, taken as a Hamiltonian function on 
$(\mathcal{H},\widetilde\omega)$ and $X_{\!\ssc\zb} (\equiv X_{\!\ssc f_{\!\zb}})$ 
the Hamiltonian vector field on $(\mathcal{P}, \omega)$. That is
in exact correspondence with Eq.(\ref{pbe}) since 
$\partial_\theta (H_{\!\ssc\zb}) =\partial_\theta (f_{\!\ssc\zb}) =0$
for any operator $\zb$. None of the vector fields has a
$\partial_r$ component, which is to be expected from the
unitary flow point of view. It may also be interested to note that
\bea
\tilde{X}_{\!\ssc\zb}^\theta
 =\frac{(1+|w|^2)}{\hbar}( w^{\tilde{m}}\partial_{{\tilde{m}}}f 
  + w^{\bar{\tilde{m}}}\partial_{\bar{\tilde{m}}}f ) 
=-(\frac{1}{z^0}\partial_{\ssc\bar 0}f+\frac{1}{\bar z^0}\partial_{\ssc 0}f) \;.
\eea
It is more convenient to focus on the covectors dual to the
Hamiltonian vector fields for which we have the expressions 
of the universal form given by the example 
$\tilde{X}_{\!\ssc\zb n}= i\partial_n f_{\!\ssc\zb}$,
{\em i.e.} components are given by the coordinate derivatives 
of Hamiltonian function multiplied by the imaginary unit $i$.
Moreover, their covariant derivatives satisfy
$\tilde{\nabla}_{\!\!m} {\tilde{X}}_{\!\ssc\zb_{\bar{n}}} 
= - \tilde{\nabla}_{\!\!{\bar{n}} } {\tilde{X}}_{\!\ssc\zb_{m}}
= -i  {\partial}_{{{m}}}  {\partial}_{\bar{{n}}}   f_{\!\ssc\zb}$
and  $\tilde{\nabla}_{\!\!m} {\tilde{X}}_{\!\ssc\zb_{{n}}} 
  = \tilde{\nabla}_{\!\!\bar{m}} {\tilde{X}}_{\!\ssc\zb_{\bar{n}}} =0$.
The form in term of the corresponding Hamiltonian 
functions is common to all K\"ahler manifolds. For the case
at hand, with the function being K\"ahlerian, the first and the
second derivatives are all the independent derivatives
 \cite{Sch,Sch2,AM},  hence their values, together with that
of the zeroth order derivative, at a point give a local
representation of the full function as the Taylor series.
The collection for all K\"ahlerian functions can be seen 
as another isomorphic description of the observable 
algebra \cite{Sch,Sch2}. Since there is such a local representation
for the algebra for each state,  we think it should be 
interpreted as the noncommutative algebra of the 
values of the observables \cite{081,079}.

\section{On the Noncommutative Coordinate Functions \label{sec4}}
So far, the coordinate systems we use on $\mathcal H$ and  $\mathcal P$
are quite generic. Now, we take a specific coordinate system
to be used for some explicit results of the basic observables 
 ${\za}_i=  \hat{x}_i + i\hat{p}_i$, $\bar\za_i = \hat{x}_i - i\hat{p}_i$.
Intuitively, these are the noncommutative coordinates of the
`phase space' for a quantum particle with the other observables 
to be seen as functions of them. Such functions will be denoted by
Greek letters, $\za$, $\zb$, $\zc$. As functions of the noncommutative 
coordinate observables, the position and momentum operators, they 
are of course operators. The latter fact readers should bear in mind 
when we talk about them as functions, and we discuss their derivatives 
with respect to the operator coordinate variables. And the term 
function is used here only in the formal sense in relation to those 
variables, without the connotation of functional values, certainly 
not one as numbers. Recall that we are talking about the observable 
algebra in the extended, mathematical, sense, as a $C^*$-algebra. 
Instead of including only the `physical observables' as Hermitian 
operators, that includes also like their complex linear combinations, 
starting from the basic `complex' coordinates variables ${\za}_i$ 
and $\bar\za_i$, which we choose to use instead of the position 
and momentum operators. All structure can be described equally 
in terms of the latter, and the simple linear relation between them 
and the `complex' ones guarantees the translation between most 
of the results put in terms of the different coordinate set, though 
some physicists would only find results in terms of $\hat{x}_i$ 
and $\hat{p}_i$ easily comprehensible. Our main task is to 
illustrate how that noncommutative `phase space' and usual 
familiar quantum phase space can be identified as the same 
symplectic geometry. 

Consider $\hat{N}_i=\frac{1}{2\hbar}\bar{\za}_i {\za}_i 
  = \frac{1}{2\hbar}  \left( \hat{x}_i^2 + \hat{p}_i^2 \right)-\frac{1}{2}$,
with $[\hat{N}_i, {\za}_j] = - \delta_{ij} {\za}_j$ and
$[\hat{N}_i, \bar{\za}_j] = \delta_{ij} \bar{\za}_j$. 
Take the simultaneous eigenstates 
$\left| n_{\!\ssc 1}, n_{\!\ssc 2},n_{\!\ssc 3}\rra$
of the number operators  $\hat{N}_i, i=1$ to 3 ({\em i.e.} satisfying
$\hat{N}_i \left| n_{\!\ssc 1}, n_{\!\ssc 2},n_{\!\ssc 3}\rra
  =n_i \left| n_{\!\ssc 1}, n_{\!\ssc 2},n_{\!\ssc 3}\rra$ for 
nonnegative integers $n_i$) as a countable orthonormal basis of 
the Hilbert space. A state is then described by the complex coordinates 
$z^{(n_{\!\ssc 1}, n_{\!\ssc 2},n_{\!\ssc 3})}$, {\em i.e.}
\bea
\left| \phi \rra=\sum_{ n_{\!\ssc 1}, n_{\!\ssc 2},n_{\!\ssc 3} }
   z^{(n_{\!\ssc 1}, n_{\!\ssc 2},n_{\!\ssc 3})} \left| n_{\!\ssc 1}, n_{\!\ssc 2},n_{\!\ssc 3}\rra \;.
\eea
The coordinates constitute a set of homogeneous coordinates 
(of the type $z^{n}$) on the projective Hilbert space from which the 
set of affine coordinates can be chosen as
$w^{(n_{\!\ssc 1}, n_{\!\ssc 2},n_{\!\ssc 3})}=
  \frac{z^{(n_{\!\ssc 1}, n_{\!\ssc 2},n_{\!\ssc 3})}}{z^{(0.0,0)}}$. 
Let us introduce the short hand index notation $[n]$ 
standing in for $(n_{\!\ssc 1}, n_{\!\ssc 2},n_{\!\ssc 3})$, 
and  further define $z^{[n]}_{i\pm} =z^{(n_{\!\ssc 1}, n_{\!\ssc 2},n_{\!\ssc 3})}_{i\pm}$,
 where  $z^{(n_{\!\ssc 1}, n_{\!\ssc 2},n_{\!\ssc 3})}_{\ssc 1\pm}
\equiv z^{(n_{\!\ssc 1}\pm {\ssc 1}, n_{\!\ssc 2},n_{\!\ssc 3})}$ \dots , etc. 
We have then the compact expressions
\bea 
 H_{\za_i}  
=\frac{1}{{2\hbar}}\sum_{[{n}]}\sqrt{{2\hbar} n_i}\bar z_{i-}^{[{n}]}z^{[{n}]}
=\frac{1}{{2\hbar}}\sum_{[{n}]}\sqrt{{2\hbar} (n_i+1)}\bar z^{[{n}]}z^{[{n}]}_{i+}
&& = \overline{H_{\bar\za_i}} \;,
\eea
and
 \bea &&
 f_{\za_i} =  \sum_{[n]}\frac{\sqrt{{2\hbar} n_i}\bar z_{i-}^{[n]}z^{[n]}}{|z|^2}
=  \sum_{[n]}\frac{\sqrt{{2\hbar} (n_i+1)}\bar z^{[n]}z^{[n]}_{i+}}{|z|^2} 
\sea \;\;
= \sum_{[n]}\frac{\sqrt{{2\hbar} n_i}\bar w_{i-}^{[n]}w^{[n]}}{1+|w|^2}
=\sum_{[n]}\frac{\sqrt{{2\hbar} (n_i+1)}\bar w^{[n]}w^{[n]}_{i+}}{1+|w|^2}
=\overline{f_{\bar\za_i}} \;.
\sea
 \eea
Note that we have used $[n]$ here including the expression
in terms of the $w^{[n]}$ coordinates where we use 
$w^{[0]} \equiv 1$. We have for the components of the covectors
\bea &&
\widetilde{X}^{\za_j}_{[{n}]}
= i\sqrt{\frac{n_j}{2\hbar}}\bar z_{[{n}]j-}
= \overline{\widetilde{{X}}^{\bar{\za}_j}_{[\bar{n}]}} \;,
\sea
\tilde{X}^{\za_j}_{[n]}
=\frac{i}{|z|^2}({\sqrt{2\hbar n_j}}\bar z_{[n]j-}-\bar z_{[n]}f_{\za_j})
= \overline{\tilde{{X}}^{\bar{\za}_j}_{[\bar{n}]}}  \;,
\sea
X^{\za_j}_{[n]}
=\frac{i}{(1+|w|^2)}({\sqrt{2\hbar n_j}}\bar w_{[n]j-}-\bar w_{[n]}f_{\za_j})
= \overline{{{X}}^{\bar{\za}_j}_{[\bar{n}]}}  
\hspace*{.2in} \qquad([n]\ne [0])  \;,
\sea
\widetilde{X}^{\za_j}_{[\bar{n}]}
= -i \sqrt{\frac{n_j+1}{2\hbar}} z_{[{n}]j+}
= \overline{\widetilde{{X}}^{\bar{\za}_j}_{[{n}]}}  \;,
\sea
\tilde{X}^{\za_j}_{[\bar n]}
=-\frac{i}{|z|^2}({\sqrt{2\hbar (n_j+1)}} z_{[n]j+} - z_{[n]}f_{\za_j})
= \overline{\tilde{{X}}^{\bar{\za}_j}_{[{n}]}}  \;,
\sea
X^{\za_j}_{[\bar n]}
=-\frac{i}{(1+|w|^2)}({\sqrt{2\hbar (n_j+1)}} w_{[n]j+} - w_{[n]}f_{\za_j})
= \overline{{{X}}^{\bar{\za}_j}_{[{n}]}}  
\qquad([n]\ne [0])  \;,
\label{cov}\eea
and the those for the vectors
\bea&&
\widetilde{X}_{\za_j}^{[n]} = -i \sqrt{\frac{2(n_j+1)}{\hbar}} {z}^{[n]}_{j+} 
= \overline{\widetilde{{X}}_{\bar{\za}_j}^{[\bar{n}]}}    \;,
\sea
\tilde{X}_{\za_j}^{[n]}
=-i \sqrt{\frac{2(n_j+1)}{\hbar}} {z}^{[n]}_{j+}  
  + i{\frac{1}{\hbar}}   {z}^{[n]} f_{\za_j}
= \overline{\tilde{{X}}_{\bar{\za}_j}^{[\bar{n}]}}  \;,
\sea
X_{\!{\za}^i}^{[n]} = -i \sqrt{\frac{2(n_j+1)}{\hbar}} {w}^{[n]}_{j+}  
  + i \sqrt{\frac{2}{\hbar}} {w}^{[0]}_{j+} {w}^{[n]}
= \overline{{{X}}_{\bar{\za}_j}^{[\bar{n}]}} 
\hspace*{.2in}
\qquad([n]\ne [0])  \;,
\sea
\widetilde{X}_{\!{\za}^i}^{[\bar{n}]} = i \sqrt{\frac{2n_j}{\hbar}} \bar{z}^{[n]}_{j-}  
= \overline{\widetilde{{X}}_{\bar{\za}_j}^{[{n}]}}   \;,
\sea
\tilde{X}_{\!{\za}^i}^{[\bar{n}]} = i \sqrt{\frac{2n_j}{\hbar}} \bar{z}^{[n]}_{j-}  
 - i{\frac{1}{\hbar}}   \bar{z}^{[n]} f_{\za_j}
= \overline{\tilde{{X}}_{\bar{\za}_j}^{[{n}]}} \;,
\sea
X_{\!{\za}^i}^{[\bar{n}]} = i \sqrt{\frac{2n_j}{\hbar}} \bar{w}^{[n]}_{j-} 
= \overline{{{X}}_{\bar{\za}_j}^{[{n}]}} 
\qquad([n]\ne [0])  \;.
\label{v}\eea
Some of the expressions here involve the undefined 
`coordinate' ${z}^{[0]}_{j-}$ which is an abuse of notation.
It always goes along with a $\sqrt{0}$ factor and the terms
vanish. The results will be useful  in analyses involving 
the noncommutative coordinates.

Let is also emphasize explicitly that the vector space of states
we are working is one of countable infinite dimensions as the 
span of the Fock states of the harmonic oscillator problem, 
as used above. In terms of the Schr\"odinger wavefunctions, 
it is the space of the rapidly decreasing functions rather than 
just square-integrable ones. $\hat{x}_i$ and $\hat{p}_i$ are 
well defined Hermitian operators, and their K\"ahlerian 
functions essentially as expectation value functions are well 
defined. Linearity gives the well defined K\"ahlerian functions 
for the $\za_i$ and $\bar\za_i$ which have been explicitly 
given. For the sake of definiteness, we can at this point take 
the observable algebra simply as the Weyl algebra of polynomials 
in $\hat{x}_i$ and $\hat{p}_i$, or  $\za_i$ and $\bar\za_i$. 
Any element of it then has the as expectation value functions 
well defined, for example through the K\"ahler  products of 
those for the coordinate observables. For full mathematical 
rigor, there are issues on topological completeness to be
concerned and the full observable algebra should include
tempered distributions instead of just the smooth functions.
However, our treatment here is considered good enough 
for the illustration of the key picture.

At the end of the last section, we mentioned that the 
full K\"ahlerian function, which is a representation of 
the corresponding operator as a quantum observable, is 
locally determined by the functional value and the values 
of the coordinate derivatives of the first two orders.  
Here above, we have given explicit results for the 
K\"ahlerian functions of the noncommutative coordinate
operators, such as $H_{\za_i}$ and $f_{\za_i}$ of the 
operator coordinates $\za_i$. Seeing $\za_i$ and $\bar\za_i$
as functions of $\hat{x}_i$ and $\hat{p}_i$ or otherwise,
they are intuitive coordinate observables. The K\"ahlerian 
functions $H_{\za_i}$ and $H_{\bar\za_i}$, or $f_{\za_i}$ 
and $f_{\bar\za_i}$ functions, are hence in a way
noncommutative coordinates for the ${\mathcal H}$ and 
${\mathcal P}$. The noncommutativity is of course to 
be seen with the K\"ahler products. We note also that 
the K\"ahlerian functions are not `symbols' of the 
corresponding operators. Though the K\"ahler product
share some similarity with the Moyal star product, they
are quite different things and not to be confused with
one another. The only place we touch on the Moyal star
product and the `symbols' is footnote 2. They are 
otherwise not used in the main text. 

The noncommutative value of an observable $\zb$ is an element 
of a noncommutative algebra each of which has a representation 
by an infinite number of complex numbers as the values of the 
independent derivatives of $f_{\!\ssc\zb}$ or $H_{\!\ssc\zb}$
 \cite{079}. Explicitly, we have
\[
[\phi] (\zb) =\{ f_{\!\ssc\zb} (\phi), {X}_{\zb_{[n]}}(\phi), {X}_{\zb_{[\bar{n}]}}(\phi), {\nabla}_{\!\!m} {{X}}_{\!\ssc\zb_{\bar{n}}}(\phi) \} \;.
\]
It can be taken, for a normalized state vector $\left|\phi \rra$
($|z|^2=\hbar$),
as
\[
[\phi] (\zb) =\{ H_{\!\ssc\zb} (\phi), \widetilde{X}_{\zb_{[n]}}(\phi), \widetilde{X}_{\zb_{[\bar{n}]}}(\phi), \widetilde{\nabla}_{\!\!m} {\widetilde{X}}_{\!\ssc\zb_{\bar{n}}}(\phi) \} \;,
\]
bearing in mind an unphysical overall phase factor ambiguity.
We use here the latter expression for simplicity. Looking at 
the noncommutative values for ${\za_i}$ and ${\bar\za_i}$, 
we have quite an amazing story. For example, the set of values 
for ${\widetilde{{X}}^{\bar{\za}_j}_{[\bar{n}]}}$, for any 
one ${\bar{\za}_j}$, is really like the full set of $z^{[n]}$ 
coordinate values. Knowing their values can completely fix 
the state vector. Of course that is based on our knowledge 
of the factors $i\sqrt{\frac{n_j}{2\hbar}}$ which are really
the constant values of the nonvanishing second derivatives, 
{\em i.e.} a $\widetilde{\nabla}_{\!\!m}
    {\widetilde{X}}_{\!\ssc\zb_{\bar{n}}}(\phi)$.
Furthermore, the simple results is special to the coordinate 
system, or basis of ${\mathcal H}$ adopted, the definition
of which involved all  ${\za_i}$ and ${\bar\za_i}$, or
$\hat{x}_i$ and $\hat{p}_i$. Hence, if we can determine 
all values of such components of a covector for a normalized 
state vector up to an overall phase, we could have all values 
of the $z^{[n]}$ coordinates up to the undetermined phase, 
and hence determine the exact physical state and all the 
information about the local representations of the full set 
of noncommutative coordinate observables, {\em i.e.} the 
Taylor series expansions of their corresponding K\"ahlerian 
functions. The calculations involved are more tedious, but 
the story is  essentially the same when we look at the set of 
${\tilde{{X}}^{\bar{\za}_j}_{[\bar{n}]}}$
or ${{{X}}^{\bar{\za}_j}_{[\bar{n}]}}$ values, except 
that we  also need the $f_{\bar{\za}_j}$ value. For example, 
we start with
$z^{[n]}= \frac{-i}{f_{\bar{\za}_j}} {\tilde{{X}}^{\bar{\za}_j}_{\ssc[\bar{n}]}}$
for all $[n]$ with $n_j=0$ and recursively each 
$z^{[n]}$ with increasing $n_j$ is given by 
$ \frac{-i}{f_{\bar{\za}_j}} ( {\tilde{{X}}^{\bar{\za}_j}_{\ssc[\bar{n}]}}
  + \sqrt{\frac{n_j}{2\hbar}} z^{[{n}]}_{j-} )$. 
So, the set of `noncommutative values' \cite{079}
for the noncommutative coordinate operators can 
be seen as carrying a lot of redundant information
about the state. The matter certainly worth further
careful studies, but here we show what we want
to focus on, that the knowing the noncommutative
values of the noncommutative coordinates in terms
of their representations as the infinite numbers of
complex numbers allows us to determine the exact 
physical state or its coordinates in the (projective) 
Hilbert space.

\section{Differential Geometric Structures for the Observable Algebra\label{sec5}}
As illustrated here in Sec.\ref{sec2} and \ref{sec3}, it is obvious 
that the commutator multiplied by $\frac{1}{i\hbar}$ is a Poisson 
bracket on the observable algebra. It is really a trivial one in 
terms of the symplectic coordinates $\hat{x}^i$ and $\hat{p}^i$,
or their complex counterparts ${\za}^i$ and $\bar{\za}^i$.
Note that these are considered coordinates for which we do not
distinguish upper and lower indices. We have then the Hamiltonian 
vector field $\mathcal{X}_{\ssc\zb}$ for ${\zb}$ given by 
\bea
\mathcal{X}_{\ssc\zb} = -\frac{1}{i\hbar} [\zb, \cdot] =-\frac{1}{i\hbar} \mbox{ad}_{\ssc\zb} \;.
\eea
Hence, it is given in terms of the adjoint action of ${\zb}$ 
on elements of the algebra, an inner derivation exactly as 
expected for the algebra  \cite{C,M}. From the geometrical 
point of view, we want to think about the Poisson bracket 
in terms of the derivatives with respect to the noncommutative 
(operator) coordinates,  
$\partial_{\hat{x}^i}$ and $\partial_{\hat{p}^i}$.  
The idea could work well with
\bea &&
\partial_{\hat{x}^i}  = \mathcal{X}_{\hat{p}^i} = -\frac{1}{i\hbar} [\hat{p}_i,\cdot] \;,
\qquad
\partial_{\hat{p}^i}= -\mathcal{X}_{\hat{x}^i} = \frac{1}{i\hbar} [\hat{x}_i,\cdot]  \;,
\eea
giving
\bea &&
\partial_i \equiv \partial_{\!{\za}^i} = -\frac{1}{2i} \mathcal{X}_{\bar{\za}^i} =
-\frac{1}{2\hbar} [\bar{\za}_i,\cdot] \;,
\qquad
 \partial_{\bar{i}} \equiv  \partial_{\!\bar{\za}^i} = \frac{1}{2i}  \mathcal{X}_{{\za}^i}=
\frac{1}{2\hbar} [\za_i,\cdot]  \;.
\eea
With the algebra  formulated for the operators $\zb$ 
as functions of the coordinate variables, $\hat{x}^i$ 
and $\hat{p}^i$ or $\za^i$ and $\bar{\za}^i$, such
differentiations can be naturally appreciated. And obviously,
the coordinates, of the real or the complex sets, are 
independent variables. In fact, the Heisenberg commutation 
relation gives $\mbox{ad}_{\hat{x}^i}$ as replacing a 
$\hat{p}^i$ factor in $\zb(\hat{p}^i,\hat{x}^i)$ by $i\hbar$ 
and $\mbox{ad}_{\hat{p}^i}$ as replacing a $\hat{x}^i$
factor by $-i\hbar$, at least for the polynomials. That is in
perfect agreement with the above expressions. The coordinate 
derivatives can be seen to be mutually commutative explicitly 
through the adjoint action. We are actually more interested 
in the expressions and results in terms of the complex 
(non-Hermitian) coordinates ${\za}^i$ and $\bar{\za}^i$ 
though we will still present some of the corresponding 
results in terms of ${\hat{x}^i}$ and $\hat{p}^i$ for
the readers' easy appreciation.

With the coordinate vector fields $\partial_{{i}}$ and  
$\partial_{\bar{i}}$, we next look at the matching 1-forms. 
The symplectic structure can be  use to give an expression 
for the differential forms  $d\zb$ as $d\zb(\mathcal{X}_\zc) 
  =\lla d\zb, \mathcal{X}_\zc \rra = \frac{1}{i\hbar}[\zb, \zc]$.
We can retrieve from that the coordinate 1-forms 
\bea &&
d\hat{x}^i (\partial_{\hat{x}^j}) 
=\delta^i_j \;,
\qquad\qquad
d\hat{x}^i (\partial_{\hat{p}^j}) =0 \;,
\sea
d\hat{p}^i (\partial_{\hat{p}^j}) =\delta^i_j \;,
\qquad\qquad
d\hat{p}^i\! (\partial_{\hat{x}^j}) =0 \;;
\eea  
as well as the matching results for  the complex 
coordinates ${\za}^i$ and $\bar{\za}^i$.  Moreover, we also
have $d\za^i (\mathcal{X}_{\ssc\zb} )=(-2i \delta^{i\bar{j}})\partial_{\bar{j}}\zb$
and  $d\bar\za^i (\mathcal{X}_{\ssc\zb} )=  (2i \delta^{\bar{i}j}) \partial_{j} \zb$,
for which a direct analog to the commutative case would
suggest seeing them as components of the $\mathcal{X}_{\ssc\zb}$
in terms of the basis of the coordinate vector fields.
We will see however that the latter idea does not work
so well in general. Similarly, we can obtain 
$d\zb (\partial_{i}) = \partial_{i} \zb$
and $d\zb (\partial_{\bar{i}}) =  \partial_{\bar{i}} \zb$,
the analog of components of the 1-forms in terms of
the coordinate 1-forms $d\za^i$ and $d\bar\za^i$.
We can go further. A symplectic structure $\Omega$
can be introduced with
\[
\Omega(d\zb,d\zc)= \{\zb,\zc\}_{\ssc \Omega}
= \Omega(\mathcal{X}_{\ssc\zb}, \mathcal{X}_{\ssc\zc} ) \;,
\]
where $\{\cdot,\cdot\}_{\ssc \Omega}$ denotes the
Poisson bracket, {\em i.e.} $\{\zb,\zc\}_{\ssc \Omega}
 = \frac{1}{i\hbar} [\zb,\zc]$. Note that 
$\mathcal{X}_{\ssc\zc} (\zb) = \{\zb,\zc\}_{\ssc\Omega}  =-\mathcal{X}_{\ssc\zb} (\zc)$.
For the coordinate vector fields or 1-forms, we can actually
write $\Omega^{i\bar{j}}\equiv \Omega(d\za^i,d\bar\za^j)=-2i \delta^{i{j}}$,
and similarly, $\Omega^{\bar{i}j}=2i\delta^{\bar{i}j}$, 
and $\Omega^{ij} =\Omega^{\bar{i}\bar{j}} =0$, as well
as $\Omega_{i\bar{j}}\equiv \Omega(\partial_i, \partial_{\bar{j}})
= \frac{i}{2} \delta_{i{j}}=- \Omega_{\bar{j}i}$ and
 $\Omega_{ij}=\Omega_{\bar{i}\bar{j}}=0$, in exact
analog to the commutative case. All the above shows 
$\hat{x}^i$ and $\hat{p}^i$ behave like three pairs
of canonical coordinates of a noncommutative 
symplectic geometry with an essentially trivial
symplectic form. 


Up to this point, we have obtained a nice picture of first 
order differential calculus on the observable algebra, or 
rather on the noncommutative geometric space behind it, 
in agreement with the general noncommutative differential 
geometry simply from  identifying the Poisson bracket with
$\frac{1}{i\hbar}[\cdot, \cdot]$ and thinking about it in 
the same way as one on a commutative symplectic manifold 
given in the differential geometric language. The structure 
may be considered dictated by the theory of quantum 
mechanics itself, and is an alternative description of the 
structure on the algebra of the $H_{\!\ssc\zb}$ functions 
or of the $f_{\!\ssc\zb}$ functions on the K\"ahler manifolds 
${\mathcal H}$ and ${\mathcal P}$, respectively. Hence,
it is natural to think about the symplectic geometry
behind the operator algebra is really the same one of
the Hilbert space or projective Hilbert space. We will look 
at that with an analysis of like a coordinate transformation 
in the next section. Note that we have not said anything 
about the metric tensors yet and therefore do not have 
the operations of lowering or raising indices, except for 
the coordinates for which there is no distinction between 
the two. The closest we get to is only the standard pairing 
between the coordinate 1-forms and the corresponding 
coordinate vector fields.  Though it is tempting to claim
an K\"ahler structure with a matching metric tensor, 
we take a special caution against that. 

To give a full differential calculus, one has to first put the 
1-forms as derivations of the 0-forms under the action of 
the differential operator $d$. Introducing the `Dirac' 
operator $D$ with
\bea \label{d2D}
d\zb  =  [D, \zb] \;,
\eea
we want $d^2=0$ and $d(\eta \eta')=d\eta\, \eta' + (-1)^k \eta \, d\eta'$ 
for forms $\eta$ and $\eta'$ with $k$ being the degree of $\eta$. 
Furthermore, one wants to extend the involution in the 
observable algebra, the complex conjugation among the $\zb$ 
functions, to all forms.  Naively extending $d=[D,\cdot]$
to all forms does not work.  The simple solution is given 
by
\bea
d\eta =    {D} \eta - (-1)^k \eta {D} 
\eea
for a $k$-form $\eta$. It is easy to see that  
$[{D},[{D}, \cdot]_{\ssc +}]  = [{D},[{D}, \cdot]]_{\ssc +} =0$,
giving $d^2=0$. It is also of interest to note that the 
Poisson bracket itself can also be naturally extended 
to the whole differential algebra \cite{CH}, with 
$d\eta =i\hbar \{{D}, \eta \}_{\ssc\Omega}$, 
by taking the graded commutator in place of the 
commutator. It is explicitly given by
\bea
 \{ \eta, \eta' \}_{\ssc\Omega}
= \frac{1}{i\hbar} \left[ \eta \eta' -  (-1)^{kk'} \eta' \eta \right] \;.
\eea
The involution, as the Hermitian conjugation of the operators, 
for the observable algebra is just our complex conjugation, 
$(\zb)^*=\bar\zb$. Obviously, $d\hat{x}^i$ and $d\hat{p}^i$,
 are to be taken as the real. $D$ then has to be taken as 
purely imaginary, {\em i.e.} ${D}^*= -D$. In addition, 
we have $(d\eta)^*= (-1)^k d(\eta)^*$. Note that 
$(\zb \zc)^*  = \bar\zc  \bar\zb$. There cannot be 
further nontrivial factors of $-1$ in the conjugate of 
product of the forms, {\em i.e.} in general we simply have 
$(\eta \eta')^*= (\eta')^* (\eta)^*$  \footnote{
Note that $(d\eta)^*= (-1)^k d(\eta)^*$ is stated as a 
convention in Ref.\cite{Ho}, also taken in Ref.\cite{C}, 
while Ref.\cite{M} adopts simply  $(d\eta)^*=  d(\eta)^*$ 
from which $(d\zb d\zc)^*= -  (d\zc)^*  (d\zb)^*$
is obtained. In our case here, there is no free choice of
convention on the matter and the latter option cannot
be taken.
}. 
We can write $\bar\eta $ for $(\eta)^*$ like in the case 
of the 0-form. In conclusion, the complex structure on our 
noncommutative space works in the usual fashion on the 
differential algebra with only the nontrivial commutation 
relations to be cared about. Or, the conjugation on products 
of the forms is exactly in line with taking all the  forms as 
operators on the same Hilbert space, though not necessarily 
as elements of the observable algebra.   However, we do 
not yet have any explicit definition of the action of $d\zb$ 
or ${D}$ on a state vector, not to say forms of the higher 
degrees, and the presented calculus remains only formal. 
That we will present in the next section.
Note that the `Dirac' operator is itself a 1-form, and an 
explicit definition of it gives explicit definitions to all forms 
as operators. 

Some checking on the forms and vector fields given above
matching well with the general notion in noncommutative
geometry is in order. We have the (Hamiltonian) vector
fields as differential operators on elements of the algebra
given as a derivations. That is quite standard, as discussed
in Ref.\cite{M} for example. Note the collection of all 
such vector fields is not a left module of the algebra, as
$\zb{\mathcal{X}}_\zc$ is in general not a derivation, and
not the Hamiltonian vector field of another element of the
algebra.  It is a free module over the center of the algebra 
though. The forms are introduced with a $d$-operator,
further expressed through the adjoint action of the 
`Dirac' operator, satisfying
$d\zb({\mathcal{X}}_\zc) ={\mathcal{X}}_\zc (\zb)$, 
and a generic $n$-form would be linear combinations
of terms in the form
\[
\zb_{\!\ssc 0} d\zb_{\!\ssc 1}  d\zb_{\!\ssc 2} \dots  d\zb_{\!\ssc n} \;,
\]
again all in line with the standard approach \cite{C,M}.
The full differential algebra of all forms then is a two-sided
module of the algebra. In particular, $d\zb \, \zc$ is
given by $d(\zb\zc) - \zb\, d\zc$. It is important to note  
the general inequality of $d\zb \, \zc\ne  \zc\, d\zb$.
For example, while $dx^2 =2xdx$ for a real variable $x$,
we have $d\zb^2 = \zb d\zb + d\zb \, \zb \ne 2\zb d\zb$
in general, which is true even for our coordinate 
operators including like $\hat{x}_i$ and $\hat{p}_i$.
So, unlike the commutative case, the coordinate 1-forms
cannot span a vector space of all 1-forms, an important
issue we will get back to below. To get something like 
a basis for all 1-forms, say if we consider just the
collection of all polynomials of $\za_i$ and $\bar\za_i$
casted back as polynomials of $\hat{x}_i$ and $\hat{p}_i$,
we would need at least all Weyl ordered monomials of the 
latter. Though we use notations like $\Omega^{i\bar{j}}$ 
above, we do not have the usual kind of expression 
of a generic $\Omega(d\zb,d\zc)$ in terms of 
$\Omega^{i\bar{j}}$ and the coordinate
derivatives as for the commutative case.  

Note that the formal 1-form $d\zb$ and $D$ operators 
will be given a more rigorous definition below through 
their action on the Hilbert space. While our picture of
the differential calculus is compatible with the general
mathematics of noncommutative geometry, our
so-called `Dirac' operator $D$ is certainly not that of
the Connes' spectral triple \cite{C}. In fact, it is not 
clear at all the operator is Dirac in the usual sense as 
in commutative geometries, and there is no spinor 
space under discussion. We use the name `Dirac' 
operator only in relation to Eq.(\ref{d2D}).

\section{Heisenberg versus Schr\"odinger Picture of the Differential 
Geometric Structures and the Transformation between 
Commutative and Noncommutative Coordinates\label{sec6}}
In the description of the time evolution in quantum mechanics,
we have the Heisenberg picture of changing observables on a fixed
state and the Schr\"odinger picture of a changing state with fixed
observables, identified by the same time dependent expectation
values. The key mathematical feature behind is  the duality 
between the observable algebra and the (geometric) manifold 
of the pure states $\mathcal{P}$. In fact, the usual Schr\"odinger 
picture is  a description completely in terms of the latter space 
when the time evolution is given in the symplectic formulation.
If that has to be clearly distinguished from the usual description
in the language of unitary flow for the state vector, one can
call it the Hamilton--Schr\"odinger picture. We will refer to it
simply as the Schr\"odinger picture in this paper and include
in it the description of $\mathcal{H}$ as a symplectic manifold.
We see that the dual descriptions, the Heisenberg and the
Schr\"odinger pictures, can be generalized to the matching 
between the structures on the observable algebra, 
with $\za_i$ and  $\bar\za_i$, or $\hat{x}_i$ and $\hat{p}_i$  
as the basic noncommutative coordinate variables on the one 
hand and the state space $\mathcal{P}$ or $\mathcal{H}$
with real/complex number symplectic coordinate variables
on the other. Duality between geometric and 
algebraic structures, exemplified in the commutative
case by a familiar finite dimensional real manifold and 
its algebra of the smooth functions, is a key perspective 
in modern mathematics along which the mathematics of 
noncommutative geometry has been developed for the
noncommutative algebras. For the specific case of the
noncommutative observable algebra in quantum mechanics
as a physical theory, our notion of Heisenberg versus 
Schr\"odinger picture descriptions gives an explicit
formulation of that duality which may be seen as 
providing a physics approach to the formulation of the
noncommutative geometry. The perspective is already
there implicitly in our description of the symplectic
differential geometric structures for the observable
algebra in the previous section. An explicit description
through the duality will be developed here. We want 
to push that to the ideal goal of such a duality map 
being fully based on a transformation between the six
noncommutative coordinates and the infinite set of the
real/complex number coordinates. 
 
We have discussed above the isomorphic picture of
 the observable algebra as an algebra of the K\"ahlerian 
functions $f_{\!\ssc\zb}$ on $\mathcal{P}$, as well as 
the corresponding $H_{\!\zb}$ functions on $\mathcal{H}$. 
The isomorphisms are essentially the duality maps. The
two are in fact only slightly different ways of writing 
the same duality map as the $f_{\!\ssc\zb}$ functions
and $H_{\!\zb}$ functions are directly related. In 
particular, their symplectic structures,  or the Poisson 
brackets, match to the same single  Poisson bracket
and the corresponding symplectic structure on the
observable algebra, as presented in the last section.
Their metric structures, however, do not. The 
implication of that has to be looked into very carefully.
From the start, each K\"ahler product algebra of the 
K\"ahlerian functions is our Schr\"odinger picture 
description of the observable algebra though, unlike in 
the classical case, the algebra is the one of a more limited 
class of functions. They generate Hamiltonian flows
which are the isometries, giving the metric a key role in the
dynamics. With the operators $\zb=\zb(\za_i, \bar\za_i)$
formulated as functions of the coordinate operators, 
the duality fits in with the intuitive idea of the position 
and momentum operators being the noncommutative
position and momentum coordinate variables of a quantum
physical/phase space, for which there is  a commutative
coordinate description requires an infinite number
of real/complex number coordinates. We will keep looking 
at both Schr\"odinger picture descriptions, on $\mathcal{P}$ 
and $\mathcal{H}$ (with the $G$ metric), in fact with
both the $z$-coordinate and $\tilde{g}$ metric and 
$w$-coordinate and ${g}$ metric for $f_{\!\ssc\zb}$ 
functions on $\mathcal{P}$, as we have been doing above.
Note that the metrics are used in two ways here, namely 
for raising and lowering indices and for the associated 
symplectic forms. We describe and discuss most of the 
results in terms of $z$-coordinates and $\tilde{g}$ metric, 
whenever explicit mathematical expressions are given. 
Direct parallel results for the other cases with similar 
properties are listed in Table~1.

\begin{table}[t]\label{table1}
 \caption{\footnotesize    
Matching results for the differential symplectic geometries under
the coordinate maps from the projective Hilbert space $\mathcal{P}$ 
and the Hilbert space $\mathcal{H}$ to the observable algebra 
as a noncommutative geometric object $\mathcal{P}_{nc}$	with 
the noncommutative (operator) coordinates. [Note that the $z^n$ 
coordinates count from $0$ while the $w^n$ coordinates from $1$ 
and   $\partial_n$ denote $\frac{\partial}{\partial z^n}$ in the 
last two columns but $\frac{\partial}{\partial w^n}$
in the $w^n$-coordinate column.]
}
\begin{center}
\begin{tabular}{||c|c|c|c|c||}    \hline
			& on $\mathcal{P}_{nc}$		& $\hat{f}^*$ to $\mathcal{P}$ ${(w)}$	& $\hat{f}^*$ to $\mathcal{P}$ ${(z)}$			&  $\hat{f}_{\!\ssc H}^*$ to $\mathcal{H}$ \\
\hline
coordinate	& 							& $w^n\!=\!\frac{z^n}{z^0},\bar{w}^n$	& $z^n,\bar{z}^n$								& $z^n,\bar{z}^n$		\\
			& ${\za}^i,\bar{\za}^i$ 		& $f_{\!\za^i}$, $f_{\!\bar{\za}^i}=\bar{f}_{\!\za^i}$  & $f_{\!\za^i}$, $f_{\!\bar{\za}^i}=\bar{f}_{\!\za^i}$ & $H_{\!\za^i}$, $H_{\!\bar{\za}^i}=\bar{H}_{\!\za^i}$\\
function		& $\zb({\za}^i,\bar{\za}^i)$	& $f_{\!\ssc\zb}(w^n,\bar{w}^n)$			& $f_{\!\ssc\zb}(z^n,\bar{z}^n)$					& $H_{\!\ssc\zb}(z^n,\bar{z}^n)$ \\
NC product		& $\zb\zc$				& $f_{\!\ssc\zb} \star_{\!\kappa} f_{\!\ssc\zc}$  & $f_{\!\ssc\zb} \star_{\!\kappa} f_{\!\ssc\zc}$ & $H_{\!\ssc\zb} \star_{\!\ssc K} H_{\!\ssc\zc}$\\

one form	&							& $dw^n,d\bar{w}^n$					& $dz^n,d\bar{z}^n$								& $dz^n,d\bar{z}^n$		  \\
			& $d\zb=[D,\zb]$			& $df_{\!\ssc\zb}= f_{\!d\ssc\zb}$		& $df_{\!\ssc\zb}= f_{\!d\ssc\zb}$				& $dH_{\!\ssc\zb}= H_{\!d\ssc\zb}$ \\
			& $d{\za}^i,d\bar{\za}^i$ 	& $df_{\!\ssc\za^i},df_{\!\ssc\bar\za^i}$	& $df_{\!\ssc\za^i},df_{\!\ssc\bar\za^i}$			& $dH_{\!\ssc\za^i},dH_{\!\ssc\bar\za^i}$\\
Poisson b.	& $\frac{1}{i\hbar} [\zb,\zc]={\Omega}(\mathcal{X}_{\ssc\zc},\mathcal{X}_{\ssc\zb})$	& $\omega ({X}_{\!\ssc\zc}, {X}_{\!\ssc\zb})$   &$\tilde\omega (\tilde{X}_{\!\ssc\zc}, \tilde{X}_{\!\ssc\zb})$				&$\widetilde{\omega} (\widetilde{X}_{\!\ssc\zc}, \widetilde{X}_{\!\ssc\zb})$
						  \\
vector field	&							& $\partial_n, \partial_{\bar{n}}$			& $\partial_n, \partial_{\bar{n}}$					& $\partial_n, \partial_{\bar{n}}$  \\
			& $\mathcal{X}_{\ssc\zb}=\frac{i}{\hbar}[\zb, \cdot]$
										& $X_{\ssc\zb}=-\{f_{\!\ssc\zb}, \cdot \}_{\!\ssc\omega}$							& $\tilde{X}_{\ssc\zb}=-\{f_{\!\ssc\zb}, \cdot \}_{\!\ssc\tilde\omega}$						& $\widetilde{X}_{\ssc\zb}=-\{H_{\!\ssc\zb}, \cdot \}_{\!\ssc\widetilde\omega}$ \\
			& $\partial_i \!=\!\left(\frac{i}{2}\right){\mathcal{X}}_{\bar\za^i}, \partial_{\bar{i}} \!=\!\left(\frac{-i}{2}\right){\mathcal{X}}_{\za^i}$    
										& $X_{\bar\za^i}, X_{\za^i}$				& $\tilde{X}_{\bar\za^i}, \tilde{X}_{\za^i}$ & $\widetilde{X}_{\bar\za^i}, \widetilde{X}_{\za^i}$\\
			&							& $J_{n}^{\;i} = \partial_{n} f_{\!\za^i}$ 	& $\tilde{J}_{n}^{\;i} = \partial_{n} f_{\!\za^i}$  &  $\widetilde{J}_{n}^{\;i} = \partial_{n} H_{\!\za^i}$\\
			&							& ${J}^{\ssc -1\,\!n}_{\,\bar{i}} = \frac{1}{2i} {X}_{\!{\za}^i}^{n}$ & $\tilde{J}^{\ssc -1\,\!n}_{\,\bar{i}} = \frac{1}{2i} \tilde{X}_{\!{\za}^i}^{n}$
      & $\widetilde{J}^{\ssc -1\,\!n}_{\,\bar{i}} = \frac{1}{2i} \widetilde{X}_{\!{\za}^i}^{n}$   \\
symplectic	& $\Omega_{\cdot\cdot}$			& ${J}^{\ssc -1}\omega_{\cdot\cdot} {J}^{\ssc -1}$   & $\tilde{J}^{\ssc -1}\tilde\omega_{\cdot\cdot} \tilde{J}^{\ssc -1}$
      & $\widetilde{J}^{\ssc -1}\widetilde\omega_{\cdot\cdot} \widetilde{J}^{\ssc -1}$\\
			& $\Omega^{\cdot\cdot}$	& ${J}\omega^{\cdot\cdot} {J}$           			& $\tilde{J}\tilde\omega^{\cdot\cdot} \tilde{J}$ & $\widetilde{J}\widetilde\omega^{\cdot\cdot} \widetilde{J}$\\
\hline
\end{tabular}\vspace*{.2in}\hrule
\end{center}
\end{table}%

Along the depicted perspective, we think of the 
K\"ahlerian functions $f_{\!\za^i}$ and $f_{\!\bar{\za}^i}$ 
as a description of the noncommutative coordinates in 
terms of the functions of the commutative coordinates, 
and $\zb$ and $f_{\!\ssc\zb}$ are the expressions of the 
same observable/function in terms of the noncommutative 
and the commutative coordinate variables, respectively. 
The six $f_{\!\za^i}$ and $f_{\!\bar{\za}^i}$ complex 
values at a point on $\mathcal{P}$ of course cannot be 
a representation of an infinite number of $z^n$ 
coordinates. They are noncommutative objects, the 
coordinates, in the K\"ahler product algebra. Their 
full noncommutative values as rather the noncommutative 
values of the operators \cite{079} serves as that 
representation. Going directly to the operators, we can 
consider an implicitly defined coordinate transformation 
map $\hat{f} : {\mathcal{P}} \to {\mathcal{P}}_{\!nc}$ 
with $\hat{f}(w^n,\bar{w}^n) = ({\za}^i,\bar{\za}^i)$,
where ${\mathcal{P}}_{\!nc}$ denotes the noncommutative
manifold of all admissible noncommutative values of
$({\za}^i,\bar{\za}^i)$. The algebraic isomorphism
sending $\zb$ to $f_{\!\ssc\zb}$ is then simply the 
pull-back transform on the corresponding functional 
space, {\em i.e.}  $f_{\!\ssc\zb}=\hat{f}^*\!({\zb})$. 
The Schr\"odinger picture of the quantum physics on 
$\mathcal{P}$ in terms of its commutative symplectic 
geometry,  is to be mapped to the Heisenberg picture 
on $\mathcal{P}_{\!nc}$ as the dual description 
based on the  noncommutative coordinate variables. 
We can also think of the map $\hat{f}$ as 
$\hat{f}(z^n,\bar{z}^n) = ({\za}^i,\bar{\za}^i)$,
with $f_{\!\ssc\zb}$ as $f_{\!\ssc\zb} (z^n,\bar{z}^n)$.
Similarly, we have 
$\hat{f}_{\!\ssc H} : {\mathcal{H}} \to {\mathcal{P}}_{\!nc}$ 
with $\hat{f}_{\!\ssc H}(z^n,\bar{z}^n) = ({\za}^i,\bar{\za}^i)$
and $H_{\!\ssc \zb} =  \hat{f}_{\!\ssc H}^*\!({\zb})$
as the algebraic isomorphism. However, we will see 
immediately below that for the coordinate maps it 
makes good sense to consider the K\"ahlerian 
functions only under the condition $|z|^2=2\hbar$, 
under which $f_{\!\ssc\zb}=H_{\!\ssc \zb}$ and 
$\hat{f}_{\!\ssc H}$ is to be taken rather as a 
${\mathcal{S}} \to {\mathcal{P}}_{\!nc}$ map. 

Firstly, we give an explicit description of differentiation 
in the Heisenberg picture, which has been given through 
the formal operator expressions in the previous section,  
 dual to the familiar counterpart in the Schr\"odinger 
picture. The two pictures of the time variation can be 
directly generalized to a generic differentiation. 
The Schr\"odinger time evolution is  given in terms of 
$z^n(t)$, for example,  with the observables
$H_{\!\ssc\zb}(t) = H_{\!\ssc\zb}(z^n(t),\bar{z}^n(t))$, 
while the Heisenberg description of the state in terms 
of the noncommutative coordinates ${\za^i}(t)$, with 
the observables  $\zb(t) = \zb(\za^i(t),\bar{\za}^i(t))$.
The two pictures are connected via 
$H_{\!\ssc\zb} (t) = H_{\!\ssc\zb(t)}$.  
For a generic infinitesimal variation, we consider
the corresponding relation 
\bea &&
\frac{1}{2\hbar} \left( \lla \delta\phi|\zb | \phi \rra
+  \lla \phi|\zb | \delta\phi \rra \right)
= dH_{\!\ssc\zb}=H_{\!\ssc d\zb}
= \frac{1}{2\hbar}  \lla \phi|d\zb | \phi \rra \;,
\label{dHb}
\eea
which can be seen as our physics definition of 
the 1-form operators $d\zb$, with 
$\left|\delta\phi \rra = \sum_n dz^n \left| n \rra$ 
as an infinitesimal state.  Equating this definition 
with the one given in Eq.(\ref{d2D}) suggests
\bea\label{Dphi}
\left|  \delta\phi \rra
=- {D} \left|\phi \rra 
\qquad  \mbox{and} \qquad
 \lla \delta\phi \right|
= \lla \phi \right| {D}\;.
\eea
Therefore, ${D}$ is antihermitian or pure imaginary,
in agreement with the results of the previous section. 
That would, however, imply $\lla \delta\phi | \phi \rra 
  = - \lla \phi |  \delta\phi \rra$ giving 
$d \lla \phi | \phi \rra =0$. The same can also be obtained 
by considering $\zb$ to be a constant operator, for 
example the identity operator. We want such $d\zb$ 
to be zero as an operator, which is not consistent with 
Eq.(\ref{dHb}) unless the $d |z|^2 =0$. In the language of the 
$\hat{f}_{\!\ssc H}$ map, for the pull-back of the 1-forms 
$d\zb$ as $\hat{f}_{\!\ssc H}^*(d\zb) =  dH_{\!\ssc\zb}$,
we surely want the pull-back of zero as a 1-form to be
zero. Moreover,  $\hat{f}^*(d\zb) =  df_{\!\ssc\zb}$ 
surely pull-backs  the zero  1-form to zero, but
 \[
df_{\!\ssc\zb} =  \frac{1}{\lla \phi|\phi \rra } \left(\lla \delta\phi|\zb | \phi \rra
  +  \lla \phi|\zb | \delta\phi \rra \right)
 -  \frac{f_{\!\ssc\zb}}{\lla \phi|\phi \rra }  (\lla \delta\phi | \phi \rra +  \lla \phi |  \delta\phi \rra ),
\]
while 
\[
f_{\!\ssc d\zb} = \frac{1}{\lla \phi|\phi \rra } \lla \phi|d\zb | \phi \rra \;.
\]
Having consistent $dH_{\!\ssc\zb}=H_{\!\ssc d\zb}$
and $df_{\!\ssc\zb}=f_{\!\ssc d\zb}$ implies 
Eq.(\ref{Dphi}) and $d |z|^2 =0$. In fact, for the 
identity operator as the constant function 
$\zb(\za^i,\bar\za^i)=1$ to be represented by the 
same constant function $H_{\!\ssc\zb}(z^n,\bar{z}^n)=1$, 
it requires $|z|^2 = 2\hbar$. So, we can have the exact 
Heisenberg and Schr\"odinger picture correspondence 
only when $H_{\!\ssc\zb}$ is taken with fixed $r=|z|$
value, most conveniently taken to be $\sqrt{2\hbar}$, 
giving the $H_{\!\ssc\zb}$ as a function on $\mathcal{S}$
and $dH_{\!\ssc\zb}$ as the 1-form. We can either use
the $z^n$ coordinates under that condition or
the set of $\{\theta, w^n, \bar{w}^n\} \; (n \ne 0)$;
note that we drop the $w^{\tilde{n}}$ notation in this 
section.  The $\hat{f}_{\!\ssc H}$ map now is to be 
taken as $\mathcal{S} \to \mathcal{P}_{\!nc}$.
However, in most of our analysis we keep the Hilbert 
space $\mathcal{H}$ as its domain and impose the 
condition $|z|^2 = 2\hbar$ in the end. Note that
$f_{\!\ssc\zb}=H_{\!\ssc\zb}$ for  $|z|^2 = 2\hbar$.

For any of the coordinate 1-forms $d\za$, we may 
write formally, following the usual case for the $\za$ 
set as if they are commutative coordinates,  
$d\za^i =  dw^{n} \hat{{J}}^{\;\;i}_{n} 
   +  dw^{\bar{n}} \hat{{J}}^{\;\;i}_{\bar{n}}$
with
$\hat{{J}}^{\;\;i}_{n} = \frac{\partial \za^i}{\partial w^{n}}$ and 
$\hat{{J}}^{\;\;i}_{\bar{n}}= \frac{\partial\za^i}{\partial w^{\bar{n}}}$.
Expressions of that kind are, however, something  
we would not quite know how to deal with. We
can look at the pull-back though, written as
\bea
\hat{f}^* (d\za^i) = dw^{n} {J}_{n}^{\;\;i}    +  dw^{\bar{n}} {J}_{\bar{n}}^{\;\;i} \;,
\eea
for which we have
\bea &&
{J}_{n}^{\;\;i}  = \frac{\partial f_{\!\za^i}}{\partial w^{n}}  =-i ({X}_{\!\za^i})_{n} \;,
\sea
{J}_{\bar{n}}^{\;\;i}  = \frac{\partial f_{\!\za^i}}{\partial \bar{w}^{n}}
=i ({X}_{\!\za^i})_{\bar{n}} \;.
\eea
A point of paramount importance to note is that
${J}_{\bar{n}}^{\;\;i}$, and its complex conjugate 
${J}_{n}^{\;\;\bar{i}}= \frac{\partial f_{\!\bar\za^i}}{\partial w^{n}}$,
are nonzero. Even $\hat{{J}}^{\;\;i}_{\bar{n}}$ 
does not look like a vanishing quantity. The coordinate map 
actually cannot be taken as a holomorphic one. That speaks 
explicitly against taking a K\"ahler structure with a trivial 
metric on $\mathcal{P}_{\!nc}$. The lack of the 
correspondence between the complex structures confirms 
that we cannot take on $\mathcal{P}_{\!nc}$ a K\"ahler 
structure connecting the symplectic and the metric one. 

For a mapping between commutative manifolds, the 
corresponding map between the tangent spaces is obtained
by pushing forward. If one follows naively that formulation,
the vector field on $\mathcal{P}_{\!nc}$, as the push-forward 
$\hat{f}_*({X})$ of the vector field ${X}$ on $\mathcal{P}$, 
should satisfy the  relation
\[
 \hat{f}_*({X}) (\zb) = {X}( \hat{f}^*(\zb)) = {X}(f_{\!\ssc \zb}) \;,
\]
and one would have
$\hat{f}_*( \frac{\partial}{\partial w^{n}})  = \hat{{J}}^{\;\;i}_{n} \partial_i
  + \hat{{J}}^{\;\;\bar{i}}_{n} \partial_{\bar{i}}$.
Again, we avoid that and look at the pull-back. In fact, 
it does not look like one can identify a vector field on 
$\mathcal{P}_{\!nc}$ as a $\hat{f}_*(\partial_{n})$, 
within the scope of our discussion in the previous 
section. To get around the difficulty, we first note that 
we should not be looking at all the functions or vector 
fields on $\mathcal{P}$. We are interested only in the 
K\"ahlerian functions as only those correspond to the 
functions of $\mathcal{P}_{\!nc}$, as pull-backs. With 
the Poisson bracket on  $\mathcal{P}_{\!nc}$, the 
noncommutative coordinate vector fields are 
Hamiltonian vector fields with the matching ones 
on $\mathcal{P}$. Therefore, we can consider each 
${X}_{\!\ssc\zb}$ as the pull-back of 
$\mathcal{{X}}_{\ssc\zb}$, and the kind of 
Hamiltonian vector fields look like the only vector fields 
we need in quantum mechanics. The coordinate vector 
field $\partial_{n}$, for example, may not have a 
push-forward on  $\mathcal{P}_{\!nc}$ as it is not 
a Hamiltonian vector field of a K\"ahlerian function. 
With the Poisson brackets discussed above, we have
\bea
{X}_{\!\ssc\zb} (\hat{f}^*(\zc))= \{f_{\!\ssc\zc},f_{\!\ssc\zb}\}_{\omega}
= \hat{f}^*\!\! \left( \{\zc,\zb\}_{\ssc\Omega} \right)
= \hat{f}^*(\mathcal{{X}}_{\ssc\zb} (\zc)) .
\eea
The last expression can be considered 
$\hat{f}^*(\mathcal{{X}}_{\ssc\zb}) (\hat{f}^* (\zc))$.
Hence, we have to give  
${X}_{\!\ssc\zb}=\hat{f}^*(\mathcal{{X}}_{\ssc\zb})$.
Then we can write 
\[
\mathcal{X}_{\ssc\zb} (\zc) 
=  \hat{f}_*({X}_{\!\ssc\zb}) (\hat{f}_*(f_{\!\ssc\zc}))
= \hat{f}_*({X}_{\!\ssc\zb}(f_{\!\ssc\zc}))
\]
and $\hat{f}_*({X}_{\!\ssc\zb}) = \mathcal{X}_{\ssc\zb}$
as the inverse. With that, we can also write the relation
\[
\hat{f}^*\!\! \left( \lla \eta, \hat{f}_*({{X}_{\!\ssc\zb}}) \rra  \right) 
  = \lla \hat{f}^*(\eta), {{X}_{\!\ssc\zb}} \rra \;,
\]
where $\eta$ is a 1-form on $\mathcal{P}_{nc}$ and
$X$ a vector field on $\mathcal{P}$. Again, 
one can check that  
\[
\lla d{\za}^i,  \mathcal{X}_{\bar\za^j} \rra 
= -2i \delta^i_j   
= \lla df_{\!\za^i}, {X}_{\!\bar\za^j} \rra 
\]
and
\[
\lla d{\za}^i,  \mathcal{X}_{\za^j} \rra = 0
= \lla df_{\!\za^i}, {X}_{\!\za^j} \rra  \;.
\]
Note that $ \lla \eta, \hat{f}_*({{X}_{\!\ssc\zb}}) \rra$ 
is in general an operator while 
$\lla \hat{f}^*(\eta), {X}_{\!\ssc\zb}\rra$ is a complex
number, which should be the pull-back of the former.
We do not have the explicit pull-back expression in 
the line above only because a constant operator (a
multiple of the identity) is pull-back by $\hat{f}^*$ 
to the same constant as a number. For the parallel 
results with $\hat{f}^*_{\!\ssc H}$ taken as the 
original $\mathcal{H} \to \mathcal{P}_{\!nc}$ 
map, however,  we have actually
$\lla dH_{\!\za^i}, \widetilde{X}_{\!\bar\za^j} \rra 
   = \frac{-i |z|^2}{\hbar} \delta^i_j$
as $\hat{f}^*_{\!\ssc H}(-2i \delta^i_j  \hat{I})$. 
Enforcing the restriction to $\mathcal{S}$, of course,
retrieves the same result as  $\hat{f}^*$. Note that 
all $\widetilde{X}_{\!\ssc\zb}$, as the vector fields on 
$\mathcal{H}$, have no components in the $\partial_r$
direction to which they are orthogonal, hence serve 
as the proper vector fields on $\mathcal{S}$. 
Similarly, using $\hat{f}$ as a map from the $z^{n}$
coordinates matched with $\tilde\omega$, the
$df_{\!\ssc\zb}$ and $\tilde{X}_{\!\ssc\zb}$ are
elements of the cotangent and tangent bundles of
$\mathcal{H}$, but are horizontal in relation to the 
vertical directions of the corresponding fiber spaces.  
The full setup for the differential symplectic  geometry 
on $(\mathcal{P}_{nc},\Omega)$ can be seen to have 
as the pull-back the differential symplectic  geometry 
on $(\mathcal{P},\omega)$, or equivalently, the one 
on $(\mathcal{P},\tilde\omega)$ or 
$(\mathcal{H},\widetilde\omega)$, which should 
be the proper way to interpret the formulation of 
the former, given in the previous section. 

Applying the pull-back to the noncommutative 
coordinate vector fields, we have  
$\hat{f}^*(\partial_i) = \frac{-1}{2i} {X}_{\!\bar{\za}^i}$
and $\hat{f}^*(\partial_{\bar{i}}) = \frac{1}{2i} {X}_{\!{\za}^i}$.
Therefore, we can write   
\bea &&
{J}^{\ssc -1\,\!n}_{\,j} 
= \frac{-1}{2i} {X}_{\!\bar{\za}^j}^{n}\;,
\qquad
{J}^{\ssc -1\,\!n}_{\,\bar{j}} 
= \frac{1}{2i} {X}_{\!{\za}^j}^{n}\;,
\eea
with
\bea
\hat{f}^*(\partial_i) = {J}^{\ssc -1\,\!n}_{\,i}  \partial_{\ssc n} 
  +  {J}^{\ssc -1\,\!\bar{n}}_{\,i}  \partial_{\ssc \bar{n}}  \;.
\eea
${J}^{\ssc -1}$ is obtained as a left inverse,
with ${J}^{\ssc -1\,\!n}_{\,j} {J}_{\ssc [{n}]}^{\;\;i} 
 + {J}^{\ssc -1\,\!\bar{n}}_{\,j} {J}_{\ssc \bar{n}}^{\;\;i} 
   =\delta^i_j$
and  ${J}^{\ssc -1\,\!n}_{\,\bar{j}} {J}_{\ssc [{n}]}^{\;\;i} 
 + {J}^{\ssc -1\,\!\bar{n}}_{\,\bar{j}} {J}_{\ssc \bar{n}}^{\;\;i}  = 0$.
We have also confirmed the last results with explicit 
expressions for the Hamiltonian vector fields and covectors
given in Eqs.(\ref{v}) and Eqs.(\ref{cov}).  The corresponding
expressions for  $\hat{f}^*_{\!\ssc H}$ similarly give 
$\widetilde{J}^{\ssc -1\,\!n}_{\,j} \widetilde{J}_{\ssc [{n}]}^{\;\;i} 
 + \widetilde{J}^{\ssc -1\,\!\bar{n}}_{\,j} \widetilde{J}_{\ssc \bar{n}}^{\;\;i} 
   =\frac{|z|^2}{2\hbar} \delta^i_j$, the pull-back of $\delta^i_j (\hat{I})$.
One can also check that, based on the pull-back relation
between the symplectic structure, we have
\bea
\hat{f}^*(\Omega_{ij}) &=&
 {{J}^{\ssc -1\,\!m}_{\,i}} {\omega}_{m\bar{n}}  {{J}^{\ssc -1\,\!\bar{n}}_{\,j}}
  +   {{J}^{\ssc -1\,\!\bar{m}}_{\,i}} {\omega}_{\bar{m}n} {{J}^{\ssc -1\,\![{n}]}_{\,j}} 
\sea
=- \frac{1}{4} \{ {f}_{\!\bar\za^i},{f}_{\!\bar\za^j}\}_{{\omega}} =0\;,
\nonumber \\
\hat{f}^*(\Omega_{i\bar{j}}) &=&
 {{J}^{\ssc -1\,\!m}_{\,i}} {\omega}_{m\bar{n}}  {{J}^{\ssc -1\,\!\bar{n}}_{\,\bar{j}}}
  +   {{J}^{\ssc -1\,\!\bar{m}}_{\,i}} {\omega}_{\bar{m}n} {{J}^{\ssc -1\,\![{n}]}_{\,\bar{j}}}
\sea
= \frac{1}{4} \{ {f}_{\!\bar\za^i},{f}_{\!\za^j}\}_{{\omega}}  = \frac{i}{2} \delta_{i\bar{j}}\;,
\label{f*Ol}\eea
and
\bea
\hat{f}^*(\Omega^{ji}) &=&
{{J}_{\ssc [{m}]}^{\;\;j}} {\omega}^{m\bar{n}} {J}_{\ssc \bar{n}}^{\;\;i}  
  + {{J}_{\ssc \bar{m}}^{\;\;j}} {\omega}^{\bar{m}n} {J}_{\ssc [{n}]}^{\;\;i}
\sea
= \{{f}_{\!\za^j},{f}_{\!\za^i}\}_{{\omega}} = 0\;,
\nonumber \\
\hat{f}^*(\Omega^{\bar{j}i}) &=&
 {{J}_{\ssc [{m}]}^{\;\;\bar{j}}} {\omega}^{m\bar{n}} {J}_{\ssc \bar{n}}^{\;\;i}  
  + {{J}_{\ssc \bar{m}}^{\;\;\bar{j}}} {\omega}^{\bar{m}n} {J}_{\ssc [{n}]}^{\;\;i}
\sea
= \{{f}_{\!\bar\za^j},{f}_{\!\za^i}\}_{\omega}= 2i \delta^{\bar{j}i} \;.
\label{f*Ou}\eea
Again, we have the corresponding results  for $(\mathcal{P},\tilde\omega)$ 
and $(\mathcal{H},\widetilde\omega)$. Expressions like 
${J}_{\ssc [{m}]}^{\;\;i}  {J}^{\ssc -1\,\![{n}]}_{\,i}
  +{J}_{\ssc [{m}]}^{\;\;\bar{i}}  {J}^{\ssc -1\,\![{n}]}_{\,\bar{i}}$
do not appear to be very sensible; and there is really no 
reason to expect otherwise. Therefore, our coordinate 
transformation  picture works perfectly well for the 
differential symplectic geometry. 

\section{Conclusions}
As physicists, we are more interested in a specific case 
that certainly has a role in our description of the nature 
rather than the general/formal mathematical considerations.
For any new mathematical ideas, we prefer to use those  
obtainable from the established physical theories and 
our related  thinking. Quantum mechanics is a very 
well established physical theory. It is also our first 
realization of the noncommutative nature of physics --- 
the noncommutativity of physical observables. The 
Heisenberg uncertainty principle clearly indicates that 
the position and momentum observables do not commute. 
Intuitively, we can understand it as the noncommutativity 
of the quantum phase space, or even the quantum model 
of the physical space. Without the modern mathematical
concept of the noncommutative geometry of an operator 
algebra, that intuitive picture can hardly be formulated. 
On the other hand, the Hilbert space seems to serve the 
purpose of describing quantum states well. A lot of 
progress in the relevant mathematics have come by 
in the recent decades and it is the time to use the 
perspectives gained there to review the theory of 
quantum mechanics, as well as to see how that first 
noncommutative physical theory informs us about 
the noncommutative geometry certainly relevant to 
our description of the nature.

The projective Hilbert space, as the manifold of pure states 
for the quantum observable algebra, is a dual object to the 
latter. The former is an infinite dimensional K\"ahler manifold 
while the latter can be thought of as a geometric structure 
with the six noncommutative coordinates. Appreciating that 
a quantum observable has the information content of infinite 
number of complex/real numbers, we seek a direct 
correspondence between the complex number 
coordinates and the noncommutative coordinates. {\em 
We use the full duality of the symplectic dynamics in the 
(Hamilton-)Schr\"odinger picture and the Heisenberg 
picture to define the differential geometry of the observable 
algebra, showing that the full differential symplectic 
geometric structures on either side match perfectly well, 
even to the extent of having an implicit coordinate map.} 
Such a map can be thought of as a coordinate transformation 
between the two as two picture of  the same geometric object. 

Our approach to noncommutative geometry is as interesting
as it is intuitive. Physicists appreciating some noncommutative 
geometry would naturally identify with the idea of the position 
and momentum observables/operators as noncommutative 
coordinates of the phase space for a quantum particle, 
though studies mostly focused on a notion of noncommutative 
spacetime as more like the configuration space. Mathematically, 
the space as the manifold of pure states has been identified as 
an object dual to the noncommutative algebra, and at least as 
one of the three candidates for its geometric picture. However, 
explicit identification and description of symplectic geometry 
of the familiar quantum phase space as a noncommutative 
geometry with the position and momentum operators as 
coordinates has not otherwise been available. In fact, without 
the new conceptual notion of their noncommutative values,
a consistent picture of the values of the six coordinates
determining a point in the projective Hilbert space 
cannot be a sensible one. Our results on the new
perspective of the coordinate transformation between
the commutative and noncommutative ones is certainly 
somewhat short of mathematical rigor. We believe the 
analysis here serves as a good first step in the direction 
for going further studies of  noncommutative geometries 
of physical interest. 

Our line of work has as one of its goal to get to an
intuitive picture of the physics of quantum mechanics
with the position and momentum operators as
geometric coordinates in the same sense as their
classical counterpart, except for the noncommutativity
 \cite{081}. We believe the present study is a good 
advance in that direction.

\begin{appendices}
\section{Appendix : On $\mathcal{P}$ from the Killing Reduction of $\mathcal{H}-\{0\}$
and the Results for the related Hamiltonian Vector Fields and Covectors}

We look at $\mathcal{H}-\{0\}$ as a complex line bundle with
$\mathcal{P}$ as the base space. We can take on it the 
conformal metric given by
\bea
\tilde{G}_{{m} {\bar{{n}}}} = \frac{2\hbar}{|z|^2} {G}_{{m} {\bar{{n}}}} \;.
\eea
It is better to replace $r$ as a coordinate by $\tau = \ln r$ 
to have $\{\tau,\theta,w^n,\bar{w}^n\}$ as the coordinate system.
$\partial_\tau = r\partial_r$ and $\partial_\theta$ are Killing
vectors the Killing reduction of which gives the Riemannian manifold 
$(\mathcal{P},\tilde{g})$ from $(\mathcal{H}-\{0\},\tilde{G})$.
Notice that we have here
\[
ds^2_{\!\ssc (\mathcal{H}=\{0\})} = 2\hbar d\tau^2 +  ds^2_{\!\ssc (\mathcal{S})} \;,
\]
which when restricted to any submanifold of constant $\tau$
gives exactly the $ds^2_{\!\ssc (\mathcal{S})}$ metric on the 
sphere $\mathcal{S}$. The metric is $\theta$-independent. A 
further projection to $\mathcal{P}$ induces on the latter the 
metric given by the $\tilde{g}$ tensor which should only 
be taken as an expression for the actual metric $g$ in the 
homogeneous coordinates, with the advantage of being globally 
applicable. The number of indices in $\tilde{g}$ is bigger 
than the dimension of the tangent space. $\tilde{g}$ can 
also be taken as a degenerate metric, $\det \tilde{g}=0$, 
on $\mathcal{H}-\{0\}$ which vanishes on the vertical part 
of the tangent space spanned by the two Killing vectors. The 
vertical tangent space is exactly the tangent space of the 
fiber manifold. A manifold with a singular Riemannian metric 
does not have an inverse metric though many of the differential 
geometric structures of the usual Riemannian manifolds may still 
be of interest \cite{K,oS}. The stationary class among such 
manifolds, which corresponds to our case at hand, has been 
a focus of mathematical studies \cite{K}. The orthogonal 
complement is the horizontal tangent space, which can be 
thought of as the tangent space for $\mathcal{P}$, or rather 
the horizontal lift of it. Horizontal tensors are defined 
accordingly. As a tensor, $\tilde{g}$ is exactly the horizontal 
lift of $g$. In the appendix of Ref.\cite{g71}, it is shown 
how horizontal tensors orthogonal to the Killing vectors and 
their covectors can be obtained from a generic tensor by 
projecting out the vertical parts. For the case at hand, we have
\bea &&
\tilde{g}_{{m} {\bar{{n}}}} = \tilde{G}_{{m} {\bar{{n}}}} 
  - [2 \partial_{\tau_{{l}}} \partial_\tau^{\,{l}}]^{-1} \partial_{\tau_{{m}}} \partial_{\tau_{\bar{{n}}}}
  - [2 \partial_{\theta_{{l}}} \partial_\theta^{\,{l}}]^{-1} \partial_{\theta_{{m}}} \partial_{\theta_{\bar{{n}}}} \;,
\sea
 \eea
where
$\partial_\tau = z^{{m}} \partial_{{m}}
  + \bar{z}^{{m}} \partial_{\bar{{m}}}$
and $\partial_\theta = iz^{{m}} \partial_{{m}}
  -i\bar{z}^{{m}} \partial_{\bar{{m}}}$. 
Notice that $\partial_{\theta_{{l}}}$ is the covector on
$(\mathcal{H}-\{0\},\tilde{G})$ here, which is different from 
the covector on $(\mathcal{H},{G})$. The structure can also 
be seen as a sub-Riemannian one  \cite{Mt}, characterized by the 
canonical 1-form  $\frac{1}{\hbar}\mbox{Im} \lla \phi | d\phi \rra
  = \frac{1}{2i\hbar}(\bar{z}_{{n}} dz^{{n}} - z_{{n}} d\bar{z}^{{n}})$,
which reduces to $-\frac{i}{\hbar}\bar{z}_{{n}} dz^{{n}}$ on $\mathcal{S}$.
The 1-form is exactly what defines the Berry connection \cite{CJ,Mt}. 
The Killing reduction technique given in Ref.\cite{g71} is a powerful
tool which, however, has apparently not been much applied in 
the mathematical studies. One can actually have an `inverse metric' 
$\tilde{g}^{{m} {\bar{{n}}}}$, for example, directly from the Killing 
reduction of $\tilde{G}^{{m} {\bar{{n}}}}$ as
 \bea &&
\tilde{g}^{{m} {\bar{{n}}}} = \tilde{G}^{{m} {\bar{{n}}}} 
  - [2 \partial_{\tau_{{l}}} \partial_\tau^{\,{l}}]^{-1} \partial_{\tau}^{\,{m}} \partial_{\tau}^{\,\bar{{n}}}
  - [2 \partial_{\theta_{{l}}} \partial_\theta^{\,{l}}]^{-1} \partial_{\theta}^{\,{m}} \partial_{\theta}^{\,\bar{{n}}} \;,
\sea
 \eea
which can and has been used above for the K\"ahler product of the 
$f_{\!\ssc\zb}$ functions. The result agrees with that obtained 
from the affine coordinates with the equivalent metric $g$ for 
$\mathcal{P}$, and that of the $H_{\!\ssc\zb}$ functions on 
$(\mathcal{H},{G})$. There is also
\bea
\tilde{g}^{{n}}_{{{m}}} 
= \delta_{{m}}^{{{n}}} - \frac{ \bar{z}_{{m}} z^{{n}}}{|z|^2}
= \tilde{g}_{{m} {\bar{{l}}}} \,\tilde{g}^{ {\bar{{l}}}{n}} 
\eea
obtainable similarly from 
$\tilde{G}_{{m}}^{{n}}=\delta_{{m}}^{{n}}$, which is
involved in the explicitly Killing reduction expressions
for the tensors. 

Since the $f_{\!\ssc\zb}$ functions, as functions on  $\mathcal{H}-\{0\}$,
are $\tau$ and $\theta$ independent, the corresponding
covectors $\partial_{{n}} f_{\!\ssc\zb}$ are naturally
horizontal, meaning they are exactly the covectors on the Killing 
reduced $\mathcal{P}$. For the covector dual to the Hamiltonian
vector field ${\tilde{X}}_{\!\ssc\zb}^{{n}}$, we have 
${\tilde{X}}_{{\!\ssc\zb}_{{n}}} \!\!
   = i  \partial_{{{n}}} f_{\!\ssc\zb}$, the same form 
as for any K\"ahler manifold with a non-degenerate metric.
It is hence exactly the same as the covector for 
${\widetilde{X}}_{\!\ssc f_{\!\ssc\zb}}$ on 
$\mathcal{H}$ as a K\"ahler manifold, and actually also 
identical to the covector for $X_{\!\ssc\zb}$. In fact, in 
terms of complex coordinates and the splits of the exterior 
derivative into holomorphic and antiholomorphic parts,
$df_{\!\ssc\zb} = \partial\! f_{\!\ssc\zb} + \bar\partial\! f_{\!\ssc\zb}$,
we always have the covector for a Hamiltonian vector field
as $i\partial\! f_{\!\ssc\zb} -i \bar\partial\! f_{\!\ssc\zb}$.
In particular, ${\tilde{X}}_{{\!\ssc\zb}}= {{X}}_{{\!\ssc\zb}}$ 
as covectors, {\em i.e.} we have  ${\tilde{X}}_{{\!\ssc\zb}}
   = {{X}}_{{\!\ssc\zb}_{\tilde{n}}} dw^{\tilde{n}}
 + {{X}}_{{\!\ssc\zb}_{\bar{\tilde{n}}}} d\bar{w}^{\tilde{n}}$ 
and the horizontal covector nicely has no component 
in the $d\tau$ (or $dr$) and $d\theta$ directions, 
though the $\{z^{{n}}, \bar{z}^{{n}} \}$ to 
$\{\tau,\theta,w^{\tilde{n}},\bar{w}^{\tilde{n}}\}$ 
transformation is not a holomorphic one.  Explicitly, 
a covector $\zeta_{{n}}$ is horizontal if 
\[ 
z^{{n}} \zeta_{{n}} = \partial_{\tau}^{\,{n}}  \zeta_{{n}} 
= (-i) \partial_{\theta}^{\,{n}}  \zeta_{{n}} =0 \;.
\]
For a horizontal vector field 
${\tilde{X}}^{{n}}$,  we have
\[
\frac{\hbar}{|z|^2} \bar{z}_{{n}} {\tilde{X}}^{{n}} =0 \;.
\]
That is satisfied by the Hamiltonian vector fields 
${\tilde{X}}_{\!\ssc\zb}$,  ${\tilde{X}}_{\!\ssc\zb}^{{n}} 
  = \tilde\omega^{{n} {\bar{{m}}}} \partial_{\bar{{m}}} f_{\!\ssc\zb}$
 (with $\tilde\omega^{{n} {\bar{{m}}}} 
   =-i \tilde{g}^{{n} {\bar{{m}}}}$).
Note that the {$w$-coordinate} derivatives $\partial_{\tilde{n}}$, 
as vectors are not horizontal, hence neither is $X_{\!\ssc\zb}$. 
That is to say, the tangent space of $\mathcal{P}$ as a manifold
is not exactly the horizontal subspace of the tangent space of 
 $\mathcal{H}-\{0\}$. ${\tilde{X}}_{\!\ssc\zb}$ is exactly
the horizontal lift of $X_{\!\ssc\zb}$, and it equals
$\frac{|z|^2}{2\hbar}{\widetilde{X}}_{\!\ssc f_{\!\ssc\zb}}$,
the Hamiltonian vector field of $f_{\!\ssc\zb}$ 
on  $(\mathcal{H},{\Omega})$. The difference is in 
the nonzero ${\tilde{X}}_{\!\ssc\zb}^\theta$.
Another expression of interest is the covariant derivative 
for the covectors. We have from the Killing reduction
\bea
\tilde{\nabla}_{\!\!{{m}}} {\tilde{X}}_{\!\ssc\zb_{\bar{{n}}}} 
=\tilde{g}^l_m \tilde{g}_{\bar{{n}}}^{\bar{{o}}} \widetilde{\nabla}_{\!\!{{l}}} {\tilde{X}}_{\!\ssc\zb_{\bar{{o}}}} \;,
\eea
where $\tilde{\nabla}_{\!\!{{m}}}$ and  $\widetilde{\nabla}_{\!\!{{l}}}$
are respectively the covariant derivatives on 
$(\mathcal{P},\tilde{g})$ and $(\mathcal{H}-\{0\},\tilde{G})$.
$\widetilde{\nabla}_{\!\!{{l}}}$ is actually quite nontrivial
 with nonvanishing Christoffel symbols given by
\bea&&
\widetilde{\Gamma}_{\!{{m}}{{n}}}^{{l}}
= -\frac{1}{2|z|^2} ( \delta_{{m}}^{{{l}}} \bar{z}_{{n}} + \delta_{{n}}^{{{l}}} \bar{z}_{{m}}) \;,
\sea
\widetilde{\Gamma}_{\!{{m}}\bar{{n}}}^{{l}}
= -\frac{1}{2|z|^2} ( \delta_{{m}}^{{{l}}} z_{{n}} - \delta_{{{m}}\bar{{n}}} z^{{l}}  ) \;,
\sea
\widetilde{\Gamma}_{\!\bar{{m}}{{n}}}^{{l}}
= -\frac{1}{2|z|^2} ( \delta_{\bar{{m}}}^{{{l}}} z_{{n}} - \delta_{{{n}}\bar{{m}}} z^{{l}}  ) \;,
\eea
and their complex conjugates as 
$\widetilde{\Gamma}_{\!\bar{{m}}\bar{{n}}}^{\bar{{l}}}$,
 $\widetilde{\Gamma}_{\!\bar{{m}}{{n}}}^{\bar{{l}}}$
and $\widetilde{\Gamma}_{\!{{m}}\bar{{n}}}^{\bar{{l}}}$.
Note that $(\mathcal{H}-\{0\},\tilde{G})$ is not a K\"ahler
manifold. The covariant derivative of the Hamiltonian 
covector, however, reduces to simply
\bea &&
\tilde{g}^l_m \tilde{g}_{\bar{{n}}}^{\bar{{o}}} {\partial}_{{{l}}} {\tilde{X}}_{\!\ssc\zb_{\bar{{o}}}} 
 = -i  {\partial}_{{{m}}}  {\partial}_{\bar{{n}}}   f_{\!\ssc\zb} \;,
\eea
which is obviously horizontal. Moreover, we have
$\tilde{\nabla}_{\!\!n} {\tilde{X}}_{\!\ssc\zb_{\bar{m}}} 
= - \tilde{\nabla}_{\!\!{\bar{m}} } {\tilde{X}}_{\!\ssc\zb_{n}}$
and  $\tilde{\nabla}_{\!\!m} {\tilde{X}}_{\!\ssc\zb_{{n}}} 
  = \tilde{\nabla}_{\!\!\bar{m}} {\tilde{X}}_{\!\ssc\zb_{\bar{n}}} =0$.
The results together with
$\tilde{\nabla}_{\!\!{{m}}} {\tilde{X}}_{\!\ssc\zb_{\bar{{n}}}} 
 =  -i  {\partial}_{{{m}}}  {\partial}_{\bar{{n}}}   f_{\!\ssc\zb}\,
( ={\partial}_{{{m}}} {\tilde{X}}_{\!\ssc\zb_{\bar{{n}}}} )$,
like ${\tilde{X}}_{{\!\ssc\zb}_{{n}}}
   = i  \partial_{{{n}}} f_{\!\ssc\zb}$, are generally valid
for K\"ahler manifolds.
 \end{appendices}

\noindent{\bf Acknowledgments \ }
We thank Suzana Bedi\'c for helping to improve the 
language of the presentation, on an earlier version of the paper.
The authors are partially supported by research grants 
number  109-2119-M-008-016
and 107-2119-M-008-011
of the MOST of Taiwan.


\begin{thebibliography}{99}

\bibitem{S}
F.W. Shultz,  Pure States as a Dual Object for $C^*$-Algebras, Commun. Math. Phys. {\bf 82}  (1982)  497-509.
\bibitem{AS}
E.M. Alfsen and F.W. Shultz, {State Spaces of Operator Algebras},
Birkh\"auser, Boston, 2001.
\bibitem{Cps}
R. Cirelli, P. Lanzavecchia, and A. Mani\`a, Normal pure states of the von Neumann algebra of bounded operators as Kahler manifold, J. Phys. A: Math. Gen. {\bf16}  (1983) 3829-3835.
\bibitem{C}
A. Connes, {Noncommutative Geometry}, Academic Press, 1994.
\bibitem{St} 
F. Stroochi,  {An Introduction to the Mathematical  Structure of Quantum Mechanics}, World Scientific, 2008.
\bibitem{Em} 
G.G.~Emch, {Algebraic Methods in Statistical Mechanics and Quantum Field Theory}. Dover, 2009.
\bibitem{kb}
X. Chen,  On a Geometric Realization of $C^*$-Algebras, Front. Math. China {\bf 9} (2014) 261-274.
\bibitem{079}
O.C.W. Kong and W.-Y. Liu, The Noncommutative Values of Quantum Observables, Chin. J. Phys. {\bf 69} (2021) 70-76. 
\bibitem{081}
O.C.W. Kong, A Geometric Picture of Quantum Mechanics with Noncommutative Values for Observables, Results Phys. {\bf 19} (2020) 103606.
\bibitem{CMP}
R.~ Cirelli, A.~ Mani\`a, and L.~ Pizzocchero, 
Quantum Mechanics as an Infinite-Dimensional Hamiltonian System with Uncertainty Structure. Part I,
J. Math. Phys. {\bf31} (1990) 2891-2897.
\bibitem{Sch} 
T.A.~Schilling,  Geometry of quantum mechanics, Ph.D dissertation,  Pennsylvania
State University, (1996). %
\bibitem{Sch2} 
A.~Ashtekar, T.A.~Schilling,  Geometrical Formulation of Quantum Mechanics, {On Einstein's Path}, A.~Harvey (ed.), Springer, New York (1998) 23-65. 
\bibitem{H}
T.W. Kibble,  Geometrization of Quantum Mechanics, Comm. Math. Phys. {\bf 65} (1979) 189-201.
\bibitem{H2}
A.~Heslot, Quantum Mechanics as a Classical Theory, Phys. Rev. D {\bf 31}, (1985) 1341-1348.
\bibitem{H3}
J.~ Anandan and Y.~ Aharonov, Geometry of quantum evolution, Phys. Rev. Lett. {\bf65} (1990) 1697-1700.
\bibitem{Ca}
V.~ Cantoni, Generalized ``Transition Probability", Commun. Math, Phys. {\bf 44} (1975) 125-128.
\bibitem{Ca2} 
V.~ Cantoni, The Riemannian Structure on the States of Quantum-like Systems, Commun. Math. Phys. {\bf56}(1977) 189-193.
\bibitem{Ca3} 
V.~ Cantoni, Superpositions of physical states:a metric viewpoint, Helv. Physica Acta {\bf 58} (1985)  956-968.
\bibitem{B}
B.A. Kupershmidt, Quantum Mechanics as an Integrable System, Phys. Lett. A  {\bf109}  (1985) 136-138.
\bibitem{B2}
A.M.~ Bloch, An Infinite-dimensional Classical Integrable System and the Heisenberg and Schr\"odinger Representations, Phys. Lett. A {\bf116} (1986) 353-355.
\bibitem{B3}
A.M.~ Bloch,  An Infinite-dimensional Hamiltonian System on Projective Hilbert Space, Trans. Amer. Math. Soc. {\bf 302} (1987) 787-796. 
\bibitem{It}
R. Cirelli, A. Mani\`a, L. Pizzocchero,  Quantum phase space formulation of Schr\"odinger mechanics, Int. J. Mod. Phys. A {\bf 6} (1991) 2133-2146.
\bibitem{spp}
R. Cirelli, M. Gatti, A. Mani\`a,  On the nonlinear extension of quantum superposition and uncertainty principles, J. Geom. Phys {\bf 29} (1999) 64-86. 
\bibitem{spp2}
R.~ Cirelli, M.~ Gatti, A.~ Mani\`a, The pure state space of quantum mechanics as Hermitian symmetric space, J. Geom. Phys {\bf45} (2003) 267-284. 
\bibitem{spp3}
A.~Corichi, Quantum superposition principle and geometry, Gen. Relat. Gravit. {\bf 38} (2006)  677-687.
\bibitem{BZ}
I. Bengtsson and K. \.Zyczkowski, Geometry of Quantum States, Cambridge University Press, Cambridge  2006.
\bibitem{CJ}
D. Chru\'sci\'nski and A. Jamio{\l}kowski, Geometric Phases in Classical and Quantum Mechanics, Birkh\"auser, Boston 2004.
\bibitem{Mt}
R. Montgomery, {A Tour of Subriemannian Geometries, Their Geodesics and Applications},
Amer. Math. Soc. 2002.
\bibitem{MS}
D. McDuff and D. Salamon,
Introduction to Symplectic Topology, 2nd ed., Clarendon Press, Oxford 1998.
\bibitem{M}
J. Madore, {An Introduction to Noncommutative Differential Geometry and its
  Physical Applications}, Cambridge University Press, Cambridge 1999.
\bibitem{CH}
C-S. Chu and P.-M. Ho, Poisson Algebra of Differential Forms, Int. J. Mod. Phys.  A {\bf 12} (1997) 5573-5587.
\bibitem{Ho}
P.-M. Ho, Riemannian Geometry on Quantum Spaces, Int. J. Mod. Phys. A {\bf 12} (1997) 923-943.
\bibitem{CHZ}
C-S. Chu, P.-M. Ho, and B. Zumino, The Quantum 2-sphere as a Complex Manifold, Z. Phys. C {\bf 70} (1996) 339-344.
\bibitem{DM}
A. Dimakis and F. Miiller-Hoissen, Quantum mechanics as non-commutative symplectic
geometry, J. Phys. A: Math. Gen. {\bf 25} (1992) 5625-5648. 
\bibitem{GP}
H. Grosse and Pre\v{s}najder, The Construction of Noncommutative Manifolds
Using Coherent States, Lett. Math. Phys. {\bf 28} (1993)  239-250.
\bibitem{066}
C.S. Chew, O.C.W. Kong, and J. Payne, A Quantum Space Behind Simple Quantum Mechanics,
Adv. High Energy Phys. \textbf{2017}  (2017) 4395918. 
\bibitem{070}
C.S. Chew, O.C.W. Kong, and J. Payne, Observables and Dynamics Quantum to Classical from a Relativity Symmetry and Noncommutative-Geometric Perspective, 
J. High Energy Phys. Gravit. Cosmol. \textbf{5} (2019) 553-586.
\bibitem{g71}
R. Geroch, A Method for Generating Solutions of Einstein's Equations, J. Math. Phys. {\bf 12} (1971) 918-924. 
\bibitem{AM}
A. Ashtekar and A. Magnon-Ashtekar, A technique for analyzing the structure of isometries, J. Math. Phys. {\bf 19} (1978) 1567-1572. 
\bibitem{K} 
D. Kupeli, Singular Semi-Riemannian Geometry, Kluwer Academic Publishers Group, Dordrecht 1996.
\bibitem{oS}
O.C.~Stoica,  On Singular Semi-Riemannian Manifolds, 	Int. J. Geom. Methods Mod. Phys. {\bf 11} (2014) 1450041.


\end{thebibliography}
\end{document}